\documentclass[oneside,reqno,12pt]{amsart}

\usepackage{rangecite}
\usepackage{epsfig}
\usepackage{euscript}

\numberwithin{equation}{section}

\newcommand{\lm}{\lambda}
\newcommand{\rh}{\rho}
\newcommand{\sg}{\sigma}
\newcommand{\bg}{\overline g}
\newcommand{\bR}{\overline R}
\newcommand{\bnabla}{\overline \nabla}
\newcommand{\tg}{\tilde g}
\newcommand{\tR}{\tilde R}
\newcommand{\tnabla}{\tilde \nabla}
\newcommand{\tG}{\tilde G}
\newcommand{\tC}{\tilde C}
\newcommand{\tzeta}{\tilde \zeta}
\newcommand{\tpi}{\tilde \pi}

\renewcommand{\mathcal}{\EuScript}

\begin{document}

\vspace*{-0.574in}
\begin{flushleft}
\scriptsize{HEP-TH/9509142, UPR-660T}
\end{flushleft}

\vspace*{0.47in}

\title[Consistent Spin-Two Coupling and Quadratic Gravitation]{Consistent 
Spin-Two Coupling and \\ Quadratic Gravitation}
\author{Ahmed Hindawi, Burt A. Ovrut, and Daniel Waldram}
\thanks{Published in Physical Review D \textbf{53} (1996), 5583--5896}
\maketitle
\vspace*{-0.3in}
\begin{center}
\small{\textit{Department of Physics, University of Pennsylvania}} \\
\small{\textit{Philadelphia, PA 19104-6396, USA}}
\end{center}

\begin{abstract}

A discussion of the field content of quadratic higher-derivative
gravitation is presented, together with a new example of a massless
spin-two field consistently coupled to gravity. The full quadratic
gravity theory is shown to be equivalent to a canonical second-order
theory of a massive scalar field, a massive spin-two symmetric
tensor field and gravity. The conditions showing that the tensor field
describes only spin-two degrees of freedom are derived. A limit of
the second-order theory provides a new example of massless spin-two
field consistently coupled to gravity. A restricted set of vacua of the
second-order theory is also discussed. It is shown that flat-space is
the only stable vacuum of this type, and that the spin-two field around flat 
space is unfortunately always ghost-like.

\end{abstract}

\thispagestyle{empty}

\renewcommand{\baselinestretch}{1.2} \large{} \normalsize{}

\vspace*{\baselineskip}

\section{Introduction}
\label{intro}

One of the simplest and most enduring ideas for extending general 
relativity is to include in the gravitational action terms which 
involve higher powers of the curvature tensor. For instance, if we
require the theory to be $\mathrm{CP}$-even and include only terms up to
quadratic in the curvature, the most general such action is
\cite{GRG-9-353}  
\begin{equation}
\label{quadS}
   S = \frac{1}{2\kappa^2} \int{ d^4x \sqrt{-g} \left[
        R + \alpha R^2 + \beta R_{\mu\nu}R^{\mu\nu} 
        + \gamma R_{\lm\mu\nu\rho}R^{\lm\mu\nu\rho} \right] },
\end{equation}
where the parameters $\alpha$, $\beta$ and $\gamma$ have dimensions of
inverse mass squared and $\kappa$ is the gravitational coupling
constant. Since the theory is purely gravitational, the parameters are
generically of the order of the inverse Planck mass squared. 
Consequently, if we expand around flat space, the quadratic terms
represent tiny, unobservable corrections to Einstein's theory. It is
of course also possible to drop the term linear in $R$ completely. However,
the pure quadratic theory, when coupled to localized positive-definite 
matter, fails to give a solution which is asymptotically flat 
\cite{CMP-2-165,PRD-10-3337,Havas-1977}. Thus to agree with Newtonian 
gravity we must include the linear term.

The quadratic theory described by \eqref{quadS} gained importance when
Stelle showed that, with $\gamma=0$, the theory is completely
renormalizable, including its coupling to matter \cite{PRD-16-953}. The
advent of string theory as a consistent theory of gravity, provided a
further motivation for studying such theories, since quadratic
corrections are explicitly present in the string effective action. In
a different direction, the quadratic theory with $\beta=\gamma=0$ has 
been used to provide a purely gravitational theory of inflation 
\cite{PLB-91-99,PLB-298-292,IJMPD-3-215}.

Here we shall be concerned with only the classical content of the
quadratic theory. The most important feature is that the equations of
motion include fourth-order derivatives of the metric, rather
than only second-order derivatives, as in Einstein's equations. The
consequence is that there are more degrees of freedom in the theory
than just the two graviton states. This can be understood by noting
that, in solving the Cauchy problem, one must now specify more
initial conditions, giving the higher-order time derivatives of
$g_{\mu\nu}$ on the initial time-slice. These new degrees of freedom
were first identified by Stelle \cite{GRG-9-353}. If we follow  his
analysis, the first observation is that classically the action
\eqref{quadS} really only describes a two parameter class of
theories. The Gauss-Bonnet combination
$\sqrt{-g}\left(R_{\lm\mu\nu\rh}R^{\lm\mu\nu\rh} 
-4R_{\mu\nu}R^{\mu\nu}+R^2\right)$ is a total divergence and so does
not contribute to the classical equations of motion. Consequently, we
can rewrite the action in terms of, for instance, only $R^2$ and the
Weyl tensor squared, $C^2=R_{\lm\mu\nu\rh}R^{\lm\mu\nu\rh}-
2R_{\mu\nu}R^{\mu\nu}+\tfrac13 R^2$, so that 
\begin{equation}
\label{R2C2S}
   S = \frac{1}{2\kappa^2} \int{ d^4x \sqrt{-g} \left[ R 
         + \frac{1}{6{m_0}^2} R^2 - \frac{1}{2{m_2}^2} C^2 \right] },
\end{equation}
where ${m_0}^{-2}=6\alpha+2\beta+2\gamma$ and
${m_2}^{-2}=-\beta-4\gamma$, and neither is restricted to be
positive. Keeping only the lowest order in the fluctuations of the
metric around flat space, Stelle showed that the theory contained, in
addition to the usual graviton, two new fields. The $R^2$ term
introduced a new scalar field with a squared-mass of ${m_0}^2$. The
$C^2$ term introduced a new spin-two field with a squared-mass of 
${m_2}^2$. However the latter field was ghost-like. This suggested
that, while the quadratic theory is renormalizable, it unfortunately
appears to be non-unitary. 

Stelle's results raise two sets of questions. The first centers around
extending his results to the complete non-linear theory. Can we identify
the new degrees of freedom in the full theory? If so, can the theory
be rewritten in a form where the new fields are made explicit and
satisfy canonical second-order equations of motion? Such a
reformulation removes the obscurity of the original higher-order
field equations, allowing for a more intuitive analysis of the theory. For
instance, we can look for the vacuum states of the theory. We can also
see if, in the full theory, the spin-two field remains ghost-like. We
note that we expect some solution of the ghost problem to exist, since
string theory, while giving an effective action which includes the
problematic quadratic terms, is nonetheless generally accepted as 
unitary. 

The second set of questions centers around the issue of coupling
spin-two fields to gravity. It has long been known that there are
problems of consistency when coupling massless fields of spin higher
than one either to themselves or to other fields
\cite{Corson53,AP-35-167,PLB-86-161,NCB-57-33,PRD-33-3613}. Even in
the massive case we may have problems with preserving
unitarity \cite{PRD-51-4543}. Yet quadratic gravity promises to be
rewritable as a theory of a spin-two field consistently coupled to a
scalar field and gravity. What is the form of the coupling? How are
the conditions required to ensure the field is pure spin-two realized?
In particular, is there a massless limit? If so, how does the theory
evade the problems of consistency which usually plague attempts to
couple a massless spin-two field to gravity? 

In this paper we will address both sets of questions. We shall give a
procedure for rewriting the full quadratic theory in a canonical
second-order form. Reducing the full action to a second-order form is
not new, though the result does not seem to be widely known. It has
been discovered and rediscovered in various places. The equivalence of
full $R+R^2$ gravity to a second-order theory of a scalar field
coupled dilatonically to gravity was first shown by Teyssandier and
Tourrenc \cite{JMP-24-2793}. Whitt reproduced and extended
these results, rewriting the second-order theory in canonical form by
making a conformal rescaling of the metric \cite{PLB-145-176}. The full
quadratic action, including a $C^2$ term, was first written in
second-order form by Magnano \textit{et al.}
\cite{GRG-19-465,CQG-5-L95,CQG-7-557} and
Jakubiec and Kijowski \cite{PRD-37-1406}. Some discussion of
the particle content of the reduced theory, though only at the
linearized level, was given by Alonso \textit{et al.}
\cite{CQG-11-865}. What is new here, however, is to put the
second-order action in a truly canonical form, separating the spin-two
and scalar degrees of freedom. This will be the main thrust of this
paper. It will provide us with a consistent theory of a  massive
spin-two field coupled to gravity. Further we shall find that the
theory has a sensible massless limit, providing an explicit example of
a completely consistent coupling of a massless higher-spin field to
gravity. We shall show how this new theory in fact falls under the
class of coupled spin-two fields discussed by Wald and Cutler 
\cite{CQG-4-1267}. 

The paper is organized as follows. In Section \ref{spin2} we discuss
the problem of coupling spin-two fields, either through
self-interaction terms or to other fields. In the massive case, the
field must satisfy a set of generalized conditions to ensure that only
the spin-two degrees of freedom propagate. In the massless case, the
problems are more extreme and imply that coupled theories are
generically inconsistent. We discuss the nature of the inconsistency
and, following \cite{PRD-33-3613} and \cite{AP-35-167}, show 
that consistency is equivalent to the presence of a new local 
symmetry. In Section \ref{R2} we discuss $R+R^2$ theory, which
introduces a new pure scalar degree of freedom. This provides us with
an example of how to rewrite quadratic gravity in a second-order form
and extract the canonical degrees of freedom.
We discuss the form and vacuum structure of the reduced theory. 
In general, by a vacuum solution we mean a stable solution 
of the second-order theory. However, in this paper we will only consider a 
restricted set of vacua in which the new scalar or spin-two degrees 
of freedom are covariantly constant. Within this context, we show 
that the only stable vacuum 
state corresponds to flat space. In Section \ref{C2} we turn to the
$R+C^2$ theory, which has a new pure spin-two degree of
freedom. Again rewriting the theory in a canonical second-order form, 
we show that suitable conditions can be derived to ensure that the new
field is indeed pure spin-two. We investigate the vacuum
structure of the reduced theory within the restricted set, 
and find that flat space is the only
stable vacuum solution and that the spin-two field fluctuations around
the vacuum remain ghost-like. In Section \ref{m2=0} we describe the
massless limit of the second-order theory given in 
Section \ref{C2}. The theory is shown to be consistent and the
associated local symmetry is identified. We find that the theory is
related to a class of theories discussed by Cutler and Wald 
\cite{CQG-4-1267,CQG-4-1279}. In Section \ref{gen2}, the general quadratic 
action 
is discussed. Reducing to a second-order form, we make the complete canonical
separation of the new scalar and spin-two degrees of freedom. Again we
discuss the vacuum structure of the theory, finding that flat space is the
only stable vacuum solution and that the spin-two field remains
ghost-like. In the final section we briefly present our conclusions. 

Throughout the paper our conventions are to use a metric of signature 
$(-+++)$ and to define the Ricci tensor as
$R_{\mu\nu}=\partial_\lm\Gamma^\lm_{\ \mu\nu}
-\partial_\mu\Gamma^\lm_{\ \lm\nu}
+\Gamma^\lm_{\ \lm\rh}\Gamma^\rh_{\ \mu\nu}
-\Gamma^\lm_{\ \rh\mu}\Gamma^\rh_{\ \lm\nu}$. 

\section{Interacting Spin-Two Theories and Consistency}
\label{spin2}

In this section we first describe the free equations of motion 
for a spin-two field and then discuss the problems of introducing
interactions in both the massive and, more critically, massless
cases. Starting in flat space, we recall that any field theory must
represent the symmetries of the spacetime, namely the Poincar\'e
group. Wigner showed that the unitary irreducible representations are
classified by the Casimir operators $M^2$ and $S^2$, which physically
represent the mass squared and the spin of the field quanta
\cite{AM-40-149}. One form of the irreducible representation with spin
two and mass $m$ is in terms of a symmetric tensor $\phi_{\mu\nu}$,
satisfying 
\begin{gather}
\label{spin2KG}
   \left( \partial^2 - m^2 \right) \phi_{\mu\nu} = 0, \\
\label{spin2conds}
   \partial^\mu \phi_{\mu\nu} = 0,  \qquad  \phi = 0,
\end{gather}
where $\phi=\eta^{\mu\nu}\phi_{\mu\nu}$. The first equation
\eqref{spin2KG} simply states that the field is an
eigenstate of the operator $M^2$ with eigenvalue $m^2$, while the
second two conditions \eqref{spin2conds} ensure that it is also an
eigenstate of $S^2$, with eigenvalue two. In fact these conditions can
be derived from an action, first given by Fierz and Pauli
\cite{HPA-12-297,PRSA-173-211,Corson53}. We write 
\begin{equation}
\label{PFS}
   S = \int{ d^4x \left[ \tfrac{1}{4} 
        \left( \partial_\mu\phi\partial^\mu\phi 
           - \partial_\mu\phi_{\nu\rh}\partial^\mu\phi^{\nu\rh}
           + 2\partial^\mu\phi_{\mu\nu}\partial^\nu\phi
           - 2\partial_\mu\phi_{\nu\rh}\partial^\rh\phi^{\nu\mu} \right)
        - \tfrac{1}{4}m^2 \left( \phi_{\mu\nu}\phi^{\mu\nu} - \phi^2 \right)
        \right] }.
\end{equation}
To understand how this action leads to the spin-two conditions
\eqref{spin2KG} and \eqref{spin2conds}, we must consider the massive
and massless cases separately. In the massive case, we start from the
corresponding equations of motion for $\phi_{\mu\nu}$, which read 
\begin{equation}
\label{PFeom}
   \partial^2\phi_{\mu\nu} 
      + \eta_{\mu\nu}\partial^\rh\partial^\sg\phi_{\rh\sg}
      - \partial_\mu\partial^\lm\phi_{\lm\nu}
      - \partial_\nu\partial^\lm\phi_{\lm\mu}
      + \partial_\mu\partial_\nu\phi - \eta_{\mu\nu}\partial^2\phi
   = m^2 \left( \phi_{\mu\nu} - \eta_{\mu\nu}\phi \right).       
\end{equation}
Taking the divergence and the trace of these equations gives a pair of 
conditions, 
\begin{equation}
\begin{aligned}
   \partial^\mu \left( \phi_{\mu\nu} - \eta_{\mu\nu}\phi \right) &= 0, \\
   2\partial^\mu\partial^\nu \left( \phi_{\mu\nu} - \eta_{\mu\nu}\phi \right)
      &= -3 m^2 \phi,
\end{aligned}
\end{equation}
respectively, which imply that
\begin{equation}
\label{PFconds}
   \partial^\mu \phi_{\mu\nu} = 0,  \qquad  \phi = 0.
\end{equation}
Here the reader should note that the divergence of the left-hand side
of the Pauli-Fierz equations \eqref{PFeom} vanishes identically, a
fact that will be important when we discuss interacting massless
spin-two fields. Substituting the conditions \eqref{PFconds} back into
the original equation of motion then gives 
\begin{equation}
   \left( \partial^2 - m^2 \right) \phi_{\mu\nu} = 0,
\end{equation}
and we find that we have succeeded in deriving the three conditions
\eqref{spin2KG} and \eqref{spin2conds}, as required. 

These conditions act to constrain the number of propagating degrees of
freedom. We start with a symmetric tensor $\phi_{\mu\nu}$ satisfying
the massive Klein-Gordon equation \eqref{spin2KG}, so that we might
expect all the ten components of $\phi_{\mu\nu}$ to
propagate. However, consider the role of the spin conditions
\eqref{spin2conds} in solving the Cauchy problem. To solve the
Klein-Gordon equations, we are required to specify twenty initial
conditions, namely each component and its time derivative on some initial  
surface. The $\phi=0$ condition implies two constraints on the initial 
conditions, since $\phi$ and $\partial_t\phi$ are initially both
zero. The divergence condition implies a further eight constraints with 
$\partial^\mu\phi_{\mu\nu}$ and $\partial_t\partial^\mu\phi_{\mu\nu}$ 
initially both set to zero. It is not clear that the latter expression
is really a constraint since it includes second-order time
derivatives. However these can be removed, since we must still satisfy
the Klein-Gordon field equations. The constraint can then be rewritten
as $\partial_i\partial^i\phi_{t\nu}-\partial_t\partial^i\phi_{i\nu}-
m^2\phi_{t\nu}=0$, where $i$ runs over the spatial indices. Thus we
are left with ten independent initial conditions, and so five degrees
of freedom. Put another way, ignoring any equations of motion, a
general symmetric tensor $\phi_{\mu\nu}$ can be decomposed into one
spin-two field, one spin-one and two scalar fields, a total of ten
degrees of freedom. The equations of motion derived from the
Pauli-Fierz action then imply two conditions \eqref{spin2conds}, which
are just sufficient to set all but the five massive spin-two degrees of
freedom to zero.

It is worth noting that the divergence condition on $\phi_{\mu\nu}$
can be promoted to the status of a conserved current. A local symmetry
can be introduced in to the Pauli-Fierz action by writing it in a form
somewhat akin to the St\"uckleberg action used for quantizing massive
gauge fields. We introduce a new vector field $\zeta_\mu$ and write 
\begin{multline}
\label{StuckS}
   S = \int{ d^4x \left[ \tfrac{1}{4} 
        \left( \partial_\mu\phi\partial^\mu\phi 
           - \partial_\mu\phi_{\nu\rh}\partial^\mu\phi^{\nu\rh}
           + 2\partial^\mu\phi_{\mu\nu}\partial^\nu\phi
           - 2\partial_\mu\phi_{\nu\rh}\partial^\rh\phi^{\nu\mu} \right) 
   \right. } \\ { \left.
        + \tfrac{1}{4}m^2 \left( 
           \left[\phi_{\mu\nu}+\partial_{(\mu}\zeta_{\nu)}\right]
           \left[\phi^{\mu\nu}+\partial^{(\mu}\zeta^{\nu)}\right]
           - \left[\phi+\partial^\mu\zeta_\mu\right]^2 \right)
        \right] }.
\end{multline}
The corresponding equations of motion are
\begin{multline}
\label{PFStuckeom}
   \partial^2\phi_{\mu\nu} 
      + \eta_{\mu\nu}\partial^\rh\partial^\sg\phi_{\rh\sg}
      - \partial_\mu\partial^\lm\phi_{\lm\nu}
      - \partial_\nu\partial^\lm\phi_{\lm\mu}
      + \partial_\mu\partial_\nu\phi - \eta_{\mu\nu}\partial^2\phi \\
      = m^2 \left[ \phi_{\mu\nu} + \partial_{(\mu}\zeta_{\nu)} 
        - \eta_{\mu\nu}\left(\phi+\partial^\lm\zeta_\lm\right) 
        \right], 
\end{multline}
\begin{equation}
\label{PFzetaeom}  
   \partial^\mu \left[ \phi_{\mu\nu} + \partial_{(\mu}\zeta_{\nu)} 
      - \eta_{\mu\nu}\left(\phi+\partial^\lm\zeta_\lm\right) \right]
      = 0.
\end{equation}
The presence of the $\zeta_\mu$ field has no effect on the dynamic
content of the theory since its field equation \eqref{PFzetaeom} is
already implied by the divergence of the $\phi_{\mu\nu}$ equation of
motion \eqref{PFStuckeom}. However, it does introduce a new gauge
symmetry. The action \eqref{StuckS} is invariant under the combined
transformation 
\begin{equation}
\label{Stucksymm}
\begin{aligned}
   \phi_{\mu\nu} &\rightarrow \phi_{\mu\nu} + \partial_{(\mu}\xi_{\nu)}, \\
   \zeta_\mu &\rightarrow \zeta_\mu - \xi_\mu.
\end{aligned}
\end{equation}
We note that the original Pauli-Fierz action \eqref{PFS} 
corresponds to the modified action \eqref{StuckS} with the gauge
choice $\zeta_\mu=0$. Further, we find that the conserved current
for the gauge symmetry \eqref{Stucksymm}, in the gauge $\zeta_\mu=0$,
is none other than $\phi_{\mu\nu}-\eta_{\mu\nu}\phi$, and thus the
divergence condition becomes a consequence of a local symmetry as
promised. We shall use this St\"uckleberg formulation in 
separating the spin-two and scalar degrees of freedom in the full
quadratic gravity theory as described in Section \ref{gen2}.

Finally we turn to the massless spin-two theory. The consequence of
setting $m=0$ is that we can no longer derive the conditions
\eqref{spin2conds} from the field equations \eqref{PFeom} following
from the Pauli-Fierz action \eqref{PFS}. However, we also find that
the action has a new local gauge symmetry. It is now invariant under the
transformation 
\begin{equation}
\label{spin2symm}
   \phi_{\mu\nu} \rightarrow \phi_{\mu\nu} + \partial_{(\mu}\xi_{\nu)}. 
\end{equation}
This immediately allows four of the ten components of $\phi_{\mu\nu}$
to be transformed away, simply by choosing a gauge. In fact, we must 
choose the gauge $\partial^\mu\phi_{\mu\nu}=0$ in order to satisfy one of 
the spin-two conditions. A residual gauge freedom, namely transformations
of the form \eqref{spin2symm} with $\partial^\mu\xi_\mu=0$,
remains. As is the case of a massless gauge field, since the gauge
$\partial^\mu\phi_{\mu\nu}=0$ implies that the remaining components of
$\phi_{\mu\nu}$ satisfy the massless Klein-Gordon equation, we find
that the residual gauge symmetry is sufficient to set $\phi$ to zero
together with three other linear combinations of the components of
$\phi_{\mu\nu}$. Thus we are left with only two propagating degrees of
freedom, the two different helicity states of a spinning, massless
field. 

Thus far we have described the content of the massive and massless
free field equations for spin-two. We would now like to consider
interactions, either through self-coupling terms or couplings to other
fields. To provide an example of the issues which arise, consider
coupling a spin-two field to gravity. The most natural approach is to
take the Pauli-Fierz action \eqref{PFS}, but to promote the ordinary
derivatives to full covariant derivatives, so that 
\begin{multline}
\label{covPFS}
   S = \int{ d^4x \sqrt{-g} \left[ \tfrac{1}{4} 
        \left( \nabla_\mu\phi\nabla^\mu\phi 
           - \nabla_\mu\phi_{\nu\rh}\nabla^\mu\phi^{\nu\rh}
           + 2\nabla^\mu\phi_{\mu\nu}\nabla^\nu\phi
           - 2\nabla_\mu\phi_{\nu\rh}\nabla^\rh\phi^{\nu\mu}\right)\right.} \\
           { \left. - \tfrac{1}{4}m^2 \left( \phi_{\mu\nu}\phi^{\mu\nu} 
            - \phi^2 \right) \right] }.
\end{multline}
The corresponding field equations are then 
\begin{equation}
\label{covPFeom}
   \nabla^2\phi_{\mu\nu} 
      + g_{\mu\nu}\nabla^\rh\nabla^\sg\phi_{\rh\sg}
      - \nabla_\mu\nabla^\lm\phi_{\lm\nu}
      - \nabla_\nu\nabla^\lm\phi_{\lm\mu}
      + \nabla_\mu\nabla_\nu\phi - g_{\mu\nu}\nabla^2\phi
   = m^2 \left( \phi_{\mu\nu} - g_{\mu\nu}\phi \right),       
\end{equation}
again with the same form as the flat-space equations, but with
ordinary derivatives replaced with covariant derivatives. Let us now
try and derive the trace and divergence conditions \eqref{spin2conds}
which in flat space restricted $\phi_{\mu\nu}$ to describe only
spin-two. Taking a trace of the field equations \eqref{covPFeom}, as
before, we obtain the condition
\begin{equation}
\label{trcoveom}
   2\nabla^\mu\nabla^\nu \left( \phi_{\mu\nu} - g_{\mu\nu}\phi \right)
      = -3 m^2 \phi. 
\end{equation}
However, when we take the divergence of the field equations new terms
arise because the covariant derivatives do not commute. We find that 
\begin{equation}
\label{divcoveom}
   R_{\nu\lm}\left(\nabla^\lm\phi - 2\nabla_\rho\phi^{\rho\lm}\right) 
       + \left(\nabla_\nu R_{\lm\rho} - 2\nabla_\lm R_{\rho\nu}
           \right) \phi^{\lm\rho}
   = m^2 \nabla^\mu \left( \phi_{\mu\nu} - g_{\mu\nu}\phi \right).
\end{equation}
Consider first the massive case. In flat space, the left-hand side of
\eqref{divcoveom} was zero, so that substituting this equation into
\eqref{trcoveom} immediately gave $\phi=0$, and then hence,
substituting back into \eqref{divcoveom}, the divergence condition 
$\partial^\mu\phi_{\mu\nu}=0$. Here this is no longer true. We note
however that a generalized form of the divergence condition does
still hold. The expressions \eqref{divcoveom} involve only first-order
time derivatives of $\phi_{\mu\nu}$, and, as such, represent four
constraint equations for $\phi_{\mu\nu}$, the same way
$\partial^\mu\phi_{\mu\nu}=0$ did in the flat-space case. We might
also hope to derive a generalized form of $\phi=0$, for instance by
taking the divergence of \eqref{divcoveom} and substituting into
\eqref{trcoveom}. However, we find that no such condition, involving
only first-order time derivatives of $\phi_{\mu\nu}$, can be found. We
are left to conclude that, though not inconsistent, the gravitationally
coupled theory no longer describes pure spin-two. In terms of the
Cauchy problem, the generalized divergence condition implies that only
six components of $\phi_{\mu\nu}$ propagate independently, but, without
a generalized trace condition, we cannot further reduce the number to
the five degrees of freedom of a massive spin-two field. 

This discussion provides us with a general prescription for
determining if an interacting theory describes pure spin-two. There
must be five conditions derivable from the equations of 
motion. If $\Phi_i$ are the other fields in a second-order theory
these conditions must be of the form
\begin{equation}
\label{genconds}
\begin{aligned}
   f_\mu (\phi_{\lm\rho},\partial_\lm\phi_{\rho\sg};
              \Phi_i,\partial_\lm\Phi_i,\partial_\lm\partial_\rho\Phi_i)
       &= 0, \\
   g (\phi_{\lm\rho},\partial_\lm\phi_{\rho\sg};
              \Phi_i,\partial_\lm\Phi_i,\partial_\lm\partial_\rho\Phi_i)
       &= 0.
\end{aligned}
\end{equation} 
The first four expressions are the generalized divergence
conditions. Linearizing in $\phi_{\mu\nu}$ and $\Phi_i$, they must
agree with the free conditions, $\partial^\mu\phi_{\mu\nu}=0$. The
final expression is the generalized trace condition, and it must read
$\phi=0$ in the linearized limit. In general, for the conditions to be
constraints on the initial conditions when solving the Cauchy problem
for the theory, they must not involve the second-order
time-derivative of $\phi_{\mu\nu}$. Here, for simplicity, we shall
assume they involve none of the second-order derivatives. 

In the massless case, the problems are more extreme. We note first
that with $m=0$ the action \eqref{covPFS} no longer has a local
symmetry. Transformations of the form \eqref{spin2symm}, with ordinary
derivatives replaced with covariant derivatives, fail because the
covariant derivatives do not commute. Critically, we also find that, with
$m=0$, the only way to satisfy the condition \eqref{divcoveom}, for any
curvature $R_{\mu\nu}$, is to set $\phi_{\mu\nu}=0$. The only
alternative is to consider \eqref{divcoveom} not as a condition on
$\phi_{\mu\nu}$, but as a condition on the curvature. This can in some
cases be a satisfactory resolution. For instance, the spin-$\tfrac32$
gravitino equation in supergravity gives a similar condition.
However, in this case, the condition is implied by the Einstein
equations and so is automatically satisfied on shell. Aragone and
Deser have investigated the possibility of a similar solution to the
problem of coupling a massless spin-two field to gravity
\cite{PLB-86-161,NCB-57-33}. However, for a large class of couplings, they
showed no such mechanism is possible. 

In fact, these problems of consistency and the loss of local gauge
invariance are related. Wald \cite{PRD-33-3613}, and earlier Ogievetsky and
Polubarinov \cite{AP-35-167}, have used this
connection as a way of deriving the possible consistent spin-two
theories. To understand this relationship, let us first, following
Wald \cite{PRD-33-3613}, define the general consistency problem. For any
massless, interacting, spin-two theory, we can separate the action
into a free part, which is the massless Pauli-Fierz action, and a part
describing the interactions,  
\begin{equation}
   S = S_{\text{PF},m=0}\left[\phi_{\mu\nu}\right] 
          + S_{\text{I}}\left[\phi_{\mu\nu},\Phi_i\right],
\end{equation}
where $\Phi_i$ are the other fields in the theory. If, for example, we
couple to gravity, the correction terms to make the ordinary
derivatives in the Pauli-Fierz action into covariant derivatives will
be included in $S_{\text{I}}$. The equations of motion following from
the general action can be written as,
\begin{equation}
\label{Ieom}
   \mathcal{T}_{\mu\nu}\left(\phi_{\rho\sg},\Phi_i\right) 
      = \frac{\delta S}{\delta\phi^{\mu\nu}}
      = {\mathcal{T}^{(1)}}_{\mu\nu} + {\mathcal{T}^{(2)}}_{\mu\nu} 
          + \dots 
      = 0,
\end{equation}
where we make an expansion such that ${\mathcal{T}^{(k)}}_{\mu\nu}$ is
$k$-th order in the fields $\phi_{\mu\nu}$ and $\Phi_i$. We know that
the first-order expression ${\mathcal{T}^{(1)}}_{\mu\nu}$ must come
from the free part of the action and so is none other than the left-hand
side of the massless Pauli-Fierz equations \eqref{PFeom}. Suppose we
now look for a perturbative solution of \eqref{Ieom}. We denote the
leading-order, linearized solution by ${\phi^{(1)}}_{\mu\nu}$. By
definition, 
\begin{equation}
   {\mathcal{T}^{(1)}}_{\mu\nu}\left({\phi^{(1)}}_{\rho\sg}\right) 
       = 0.
\end{equation}
We can similarly solve for ${\Phi^{(1)}}_i$, the linearized solution
of the $\Phi_i$ equations of motion. Now consider solving for the
quadratic corrections to ${\phi^{(1)}}_{\mu\nu}$, denoted
${\phi^{(2)}}_{\mu\nu}$. They must satisfy
\begin{equation}
\label{Ieom2}
   {\mathcal{T}^{(1)}}_{\mu\nu}\left({\phi^{(2)}}_{\rho\sg}\right) 
      + {\mathcal{T}^{(2)}}_{\mu\nu}
            \left({\phi^{(1)}}_{\rho\sg},{\Phi^{(1)}}_i\right) 
      = 0.
\end{equation}
As we noted earlier $\partial^\mu{\mathcal{T}^{(1)}}_{\mu\nu}
\left(\phi_{\rho\sg}\right)$ is identically zero for any
$\phi_{\rho\sg}$. Thus, if we take the divergence of equation
\eqref{Ieom2}, we are left with the relation 
\begin{equation}
\label{divIeom2}
   \partial^\mu{\mathcal{T}^{(2)}}_{\mu\nu}
       \left({\phi^{(1)}}_{\rho\sg},{\Phi^{(1)}}_i\right) = 0. 
\end{equation}
This implies a condition on ${\phi^{(1)}}_{\mu\nu}$ and
${\Phi^{(1)}}_i$. The result is that not all the linearized solutions
are necessarily consistent with the equation for the solution to the
next order in the perturbation expansion. As we go to higher orders in
the expansion, we shall find more and more conditions. Indeed it may
be that none of the linearized solutions are compatible with these
conditions, or as we found in the case of the naive coupling of a
spin-two field to gravity only the trivial solution $\phi_{\mu\nu}=0$
survives. This lack of a sensible procedure for perturbatively solving
the field equations is one way of defining the problem of inconsistency. 

To avoid inconsistency, we require \eqref{divIeom2} to be satisfied by
any solution of the linearized equations, and that the similar conditions,
arising at higher-orders in the perturbation expansion, are also
automatically satisfied by solving the field equations to one order
lower in the expansion. One way of ensuring this is if the full
interacting action has a local symmetry of a particular form. For a
second-order theory, we require the action to be invariant under
infinitesimal transformations of the form 
\cite{AP-35-167,PRD-33-3613,CQG-4-1267}
\begin{equation}
\label{gensymm}
\begin{aligned}
   \delta\phi_{\mu\nu} &= \partial_{(\mu}\xi_{\nu)} 
        + {\alpha_{\mu\nu}}^{\rho\sg} \partial_{(\rho}\xi_{\sg)}
        + {\beta_{\mu\nu}}^\rho \xi_\rho, \\
   \delta\Phi_i &= {\gamma_i}^{\rho\sg} \partial_\rho\xi_\sg 
        + {\epsilon_i}^\rho \xi_\rho. 
\end{aligned}
\end{equation}
Here the coefficients ${\alpha_{\mu\nu}}^{\rho\sg}$ and 
${\gamma_i}^{\rho\sg}$ are taken to be general functions of
$\phi_{\mu\nu}$ and $\Phi_i$, while ${\beta_{\mu\nu}}^\rho$ and
${\epsilon_i}^\rho$ may also depend on $\partial_\lm\phi_{\mu\nu}$ and 
$\partial_\lm\Phi_i$. When linearizing the transformation in
$\phi_{\mu\nu}$ and $\Phi_i$, all the coefficients must vanish, so
that  we return to the old symmetry of the massless Pauli-Fierz
action. If one calculates the corresponding conserved current, as is
done in Wald \cite{PRD-33-3613} and Ogievetsky and Polubarinov
\cite{AP-35-167}, one finds that, order by order in the
perturbation expansion, the divergence conditions, such as
\eqref{divIeom2}, are satisfied and there is no consistency problem. We
also note that, since the symmetry has the same form as the gauge
symmetry of the free action, we expect to be able to eliminate all but
two propagating components of $\phi_{\mu\nu}$, so that it truly
describes a massless spin-two field.

Both Ogievetsky and Polubarinov \cite{AP-35-167} and
Wald \cite{PRD-33-3613} have used this formalism to prove that a generally
covariant theory, such as Einstein gravity, where the $\phi_{\mu\nu}$
field really describes the metric of a curved spacetime, is the only
consistent theory of a single spin-two field. Cutler and Wald
\cite{CQG-4-1267} also considered consistent collections of spin-two
fields. This will include any consistent theory of a massless spin-two
field coupled to gravity. Wald \cite{CQG-4-1279} obtained the very
interesting result that such theories correspond to theories of a
metric on algebra-valued manifolds, where the coordinates are elements
of an associative, commutative algebra. Here, we shall not be
concerned with this geometrical interpretation, but it will be
important to note Wald's result that at least one of the spin-two
fields in a consistent theory must be ghost-like. We shall return to
this issue in Section \ref{m2=0}. 

In conclusion, for a massive, interacting, symmetric tensor field, we 
require a set of conditions \eqref{genconds} to hold if the field is to
describe only spin-two. In the massless case, we require the action to
have a local symmetry of the form \eqref{gensymm} in order for the
theory to be consistent and describe spin-two. These are the
conditions we must show apply when we attempt to extract the spin-two
degree of freedom from quadratic gravity.

\section{The $R + R^2$ Theory}
\label{R2}

Before turning to general quadratic gravity, let us first consider the
special case where we include only the scalar curvature quadratic
correction, so that 
\begin{equation}
\label{R2S}
   S = \frac{1}{2\kappa^2} \int{ d^4x \sqrt{-g} 
        \left( R + \frac{1}{6m^2} R^2 \right) }. 
\end{equation}
From Stelle's linearized calculations, we expect the $R^2$ term to
introduce a new scalar degree of freedom into the theory. Extracting
this degree of freedom, by rewriting the theory in a canonical
second-order form, will provide us with a simple example of a
procedure we shall use throughout this paper. Having obtained the
reduced, second-order theory, we shall briefly discuss its physical
content, in particular the vacuum solutions. 

In rewriting the theory in a canonical second-order form, we will
follow Whitt \cite{PLB-145-176}. The procedure is first to 
introduce an auxiliary field for the scalar curvature $R$, thus giving 
an action with second-order equations 
of motion. This is the analog of forming the Helmholtz Lagrangian, which 
reduces a second-order theory to a first-order theory in terms of field 
variables and conjugate momenta. Here we instead reduce a fourth-order 
theory to second-order, with the auxiliary field playing the role of the 
conjugate momentum. Then, by redefining the fields to grow an explicit
kinetic energy term for the new field, the action is transformed to
the canonical form for a scalar field coupled to Einstein gravity. 

Thus, introducing a dimensionless auxiliary field $\lm$, the action 
\eqref{R2S} can be rewritten as 
\begin{align}
   S &= \frac{1}{2\kappa^2} \int{ d^4x \sqrt{-g} \left[ 
       R + \frac{1}{6m^2} R^2 
       - \frac{1}{6m^2} \left(R-3m^2\lm\right)^2 \right] } \notag \\
     &= \frac{1}{2\kappa^2} \int{ d^4x \sqrt{-g} \left[
       \left(1+\lm\right) R - \tfrac{3}{2} m^2 \lm^2 \right] }.
\label{R2auxS}
\end{align}
The $\lm$ equation of motion is algebraic, giving
$R=3m^2\lm$. Substituting this solution back into the action clearly returns
one to the original higher-derivative form and so the theories are
equivalent. The equations of motion for $g_{\mu\nu}$ and $\lm$ are now
second-order, but the action is not yet in a canonical form. With this
in mind, consider the transformation of the Ricci scalar under a
conformal rescaling of the metric. If $\bg_{\mu\nu}=e^\chi
g_{\mu\nu}$, in terms of the new metric and the covariant derivative
$\bnabla_\lm \bg_{\mu\nu}=0$, we have 
\begin{equation}
   \sqrt{-g}R = \sqrt{-\bg}e^{-\chi} \left( \bR + 3\bnabla^2\chi 
       - \tfrac{3}{2}\left(\bnabla\chi\right)^2 \right).
\end{equation}
Comparing with \eqref{R2auxS} we see that by choosing 
$\chi=\log\left(1+\lm\right)$, and conformally rescaling the metric we 
can produce a canonical Einstein-Hilbert term. Furthermore we also generate 
a canonical kinetic energy term. Thus, transforming from
$(g_{\mu\nu},\lm)$ to $(\bg_{\mu\nu},\chi)$ and dropping total
derivative terms, the action becomes 
\begin{equation}
\label{canR2}
   S = \frac{1}{2\kappa^2} \int{ d^4x \sqrt{-\bg} \left[
        \bR - \tfrac{3}{2}\left(\bnabla\chi\right)^2 
        - \tfrac{3}{2}m^2\left(1-e^{-\chi}\right)^2 \right] }. 
\end{equation}
This is the canonical form for Einstein gravity coupled to a new scalar 
field $\chi$ and reproduces the result derived by Whitt \cite{PLB-145-176}. 
One subtlety here is that $\chi$ is only defined as a real field 
for $\lm>-1$. Note, however, that despite this restriction on $\lm$, the 
range of $\chi$ is unrestricted with $-\infty<\chi<\infty$. For
$\lm<-1$ we must introduce, instead, the real field
$\chi=\log\left(-1-\lm\right)$, the effect of which is to change the
sign of the overall normalization of the action. At the one special
point $\lm=-1$, the conformal rescaling becomes degenerate. Thus at
this point the action cannot be put in canonical form, although there
is no problem with the non-canonical formulation in terms of
$\lm$. Here, and throughout this paper, we will restrict our attention
to the range of the auxiliary field(s) for which the transformed
action has, when possible, the usual normalization. Finally, we also
note that strictly the action \eqref{canR2} is not quite in canonical
form since the scalar field $\chi$ is dimensionless and has a
non-canonical normalization of its kinetic energy. This is, of course,
easily remedied by introducing a dimensionful rescaled field
$\chi'=(3/2\kappa^2)^{1/2}\chi$. For simplicity, throughout this paper,
we will not make this final trivial rescaling. 

Clearly the scalar field kinetic energy in action \eqref{canR2} has the 
usual sign, and, therefore, is not a ghost. The $\bg_{\mu\nu}$ and
$\chi$ equations of motion following from \eqref{canR2} have the
conventional forms 
\begin{gather}
\label{canR2geom}
   \bR_{\mu\nu} - \tfrac12\bg_{\mu\nu}\bR = 
         \tfrac32 \left[ \bnabla_\mu\chi\bnabla_\nu\chi
             - \tfrac12 \bg_{\mu\nu} \left(\bnabla\chi\right)^2 \right] - 
         \tfrac32\bg_{\mu\nu}V(\chi), \\
\label{canR2chieom}
   \bnabla^2\chi = \frac{dV}{d\chi},
\end{gather}
respectively, where here the potential is given by  
\begin{equation}
\label{canR2V}
    V(\chi)= \tfrac12 m^2\left(1-e^{-\chi}\right)^2.
\end{equation}
Consider the vacuum solutions of these equations. If we restrict
ourselves to solutions where the scalar field is covariantly constant,
\begin{equation}
\label{covconst}
   \bnabla_\mu\chi = \partial_\mu\chi = 0,
\end{equation}
it follows from \eqref{canR2chieom}, that, $\chi$ must extremize that
potential. The $\bg_{\mu\nu}$ equation of motion \eqref{canR2geom} then
implies that the spacetime is a space of constant curvature, a
deSitter or anti-deSitter solution, with 
\begin{equation}
\label{R2vac}
   \bR = 6V(\chi). 
\end{equation}
Since the original metric $g_{\mu\nu}$ is related to $\bg_{\mu\nu}$ by
a conformal rescaling, $g_{\mu\nu}=e^{-\chi}\bg_{\mu\nu}$,and $\chi$
is constant, it follows that 
\begin{equation}
   R = e^\chi \bR.
\end{equation}
Thus these vacuum states are also spaces of constant curvature with
respect to the original metric, although the radius of curvature is in
general different. 

\begin{figure}[ht]
   \centerline{\psfig{figure=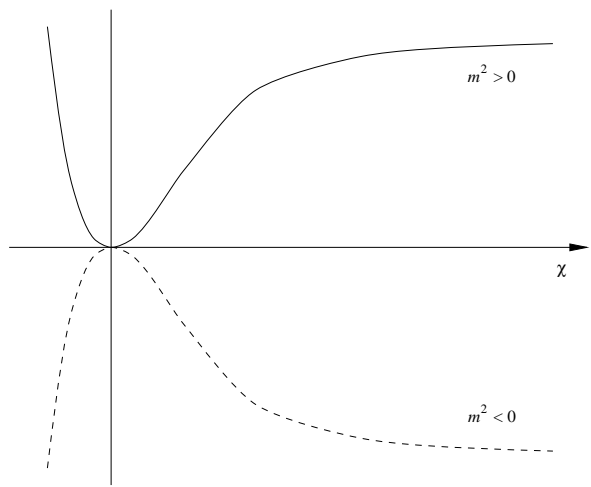,height=2.5in}}
   \caption{$V(\chi)$ for quadratic higher-derivative gravity}
   \label{fig:R2V}
\end{figure}

The potential energy \eqref{canR2V} is plotted in Figure~\ref{fig:R2V}. 
For $m^2>0$ the potential has a single stable minimum at $\chi=0$, with scalar
excitations of mass $m$, which we recall is generically of order of
the Planck mass. For $m^2<0$ we have no stable minimum. Therefore, the
only stable vacuum solution is for $m^2>0$ and $\chi=0$. Since
$V(0)=0$, it follows that $\bR=R=0$ and, hence, that
$\bg_{\mu\nu}=g_{\mu\nu}=\eta_{\mu\nu}$. That is, the only stable
vacuum solution where the scalar field is constant is trivial flat space with 
vanishing scalar vacuum expectation value. 

In closing this section it is worth emphasizing two facts. The first is 
that it is clear from this discussion that the quadratic gravity action 
\eqref{R2S} which involves only the metric field $g_{\mu\nu}$, 
in the range of the scalar curvature given by $R>-3m^2$, is completely 
equivalent to the action \eqref{canR2} which involves not only a metric field 
$\bg_{\mu\nu}$ described by the usual Einstein theory, but 
also a massive, non-ghost scalar field $\chi$. The conclusion is that the 
modification of Einstein's theory obtained by adding the $R^2$ term in 
\eqref{R2S}, increases the number of propagating degrees of freedom contained 
in $g_{\mu\nu}$ from the two helicity states of the usual graviton to three. 
This can easily be seen by considering the fourth-order $g_{\mu\nu}$ 
equations of motion associated with action \eqref{R2S}. The formalism
we have used here, introducing the auxiliary field $\lm$ and
transforming to the $(\bg_{\mu\nu},\chi)$ variables, simply makes this
fact manifest. We want to emphasize that the new degree of freedom is
physical and cannot be removed from the theory by, for example, a
field redefinition or gauge transformation. 

The second fact we would like to point out is that the new scalar degree of 
freedom has a potential energy that is uniquely determined by the original 
action \eqref{R2S}. This opens the possibility that $\chi$ or, equivalently, 
the original metric $g_{\mu\nu}$ might have a non-trivial vacuum state. We 
see from Figure~\ref{fig:R2V} that this is not the case. However, 
it is possible that more general higher-derivative gravity theories do have 
non-trivial gravitational vacua. Indeed, if one considers actions
which are general functions of the curvature, we have shown that this
is the case \cite{PRD-53-5597}. 

\section{The $R + C^2$ Theory}
\label{C2}

We now turn to the $R+C^2$ quadratic theory, which from Stelle's
linearized analysis we expect to introduce new pure spin-two degrees
of freedom. The action is given by 
\begin{equation}
   S = \frac{1}{2\kappa^2} \int{ d^4x \sqrt{-g} \left[
        R - \frac{1}{2m^2}C^2 \right] }.
\end{equation}
We would like to introduce an auxiliary field to reduce the theory to
a second-order form as we did in the case of the $R+R^2$ theory. First
though, it is useful to rewrite the action in terms of only the Ricci
tensor $R_{\mu\nu}$. This can be done by extracting a total
divergence, a Gauss-Bonnet term, which will not contribute to the
classical equations of motion. We have,
\begin{align}
   S &= \frac{1}{2\kappa^2} \int{ d^4x \sqrt{-g} \left[
           R - \frac1{2m^2} C^2 \right] } \notag \\
     &= \frac{1}{2\kappa^2} \int{ d^4x \sqrt{-g} \left[
           R - \frac1{2m^2} \left( R_{\lm\mu\nu\rh}R^{\lm\mu\nu\rh} 
               - 2R_{\mu\nu}R^{\mu\nu} + \tfrac13 R^2 \right)
           \right] }  \notag \\
     &= \frac{1}{2\kappa^2} \int{ d^4x \sqrt{-g} \left[
           R - \frac1{m^2} \left( R_{\mu\nu}R^{\mu\nu} - \tfrac13 R^2 \right) 
           - \frac1{2m^2} \left( R_{\lm\mu\nu\rh}R^{\lm\mu\nu\rh} 
               - 4R_{\mu\nu}R^{\mu\nu} + R^2\right)
           \right] } \notag \\
     &= \frac{1}{2\kappa^2} \int{ d^4x \sqrt{-g} \left[
           R - \frac1{m^2} \left( R_{\mu\nu}R^{\mu\nu} - \tfrac13 R^2 \right) 
           \right] }. 
\label{Ricci2S}
\end{align}
Written in the form given in the last line, we can now reduce the
action to a second-order, but not canonical, form by introducing a
symmetric tensor auxiliary field $\pi_{\mu\nu}$. Such a procedure was
first given by Magnano \textit{et al.} \cite{GRG-19-465,CQG-5-L95,CQG-7-557} and
Jakubiec and Kijowski \cite{PRD-37-1406}. We shall use a
slightly different formulation in which it is easier to demonstrate
that the new field $\pi_{\mu\nu}$ is pure spin-two. We write 
\begin{align}
   S &= \frac{1}{2\kappa^2} \int{ d^4x \sqrt{-g} \left[
       R - \frac{1}{2m^2} C^2 \right] } \notag \\
     &= \frac{1}{2\kappa^2} \int{ d^4x \sqrt{-g} \left[
       R - \frac{1}{m^2} \left( R_{\mu\nu}R^{\mu\nu} - \tfrac{1}{3}R^2 \right) 
       \right] } \notag \\
     &= \frac{1}{2\kappa^2} \int{ d^4x \sqrt{-g} \left[
       R - G_{\mu\nu}\pi^{\mu\nu} 
       + \tfrac{1}{4}m^2 \left( \pi_{\mu\nu}\pi^{\mu\nu} - \pi^2 \right)
       \right] }, 
\label{C2auxS}
\end{align}
where $\pi=\pi_{\mu\nu}g^{\mu\nu}$ and 
$G_{\mu\nu}=R_{\mu\nu}-\tfrac{1}{2}g_{\mu\nu}R$. The auxiliary field equation 
of motion then is
\begin{equation}
\label{C2pieom}
   G_{\mu\nu} = \tfrac{1}{2}m^2 \left( \pi_{\mu\nu} - g_{\mu\nu}\pi \right),
\end{equation}
which gives $\pi_{\mu\nu}=2{m}^{-2}\left(R_{\mu\nu}-\frac{1}{6}
g_{\mu\nu}R\right)$. Substituting this expression back into the action yields 
the original fourth-order theory.

The metric equation of motion for the reduced action given in the last
line of \eqref{C2auxS} reads
\begin{multline}
\label{C2geom}
   \nabla^2\pi_{\mu\nu} + g_{\mu\nu}\nabla^\rh\nabla^\sg\pi_{\rh\sg}
      - \nabla_\mu\nabla^\lm\pi_{\lm\nu} - \nabla_\nu\nabla^\lm\pi_{\lm\mu}
      + \nabla_\mu\nabla_\nu\pi - g_{\mu\nu}\nabla^2\pi \\
      + R_\mu^{\ \ \lm} \left(\pi_{\lm\nu}-\tfrac{1}{2}g_{\lm\nu}\pi\right)
      + R_\nu^{\ \ \lm} \left(\pi_{\lm\mu}-\tfrac{1}{2}g_{\lm\mu}\pi\right)
      - \tfrac{1}{2}g_{\mu\nu}R^{\rh\sg}
             \left(\pi_{\rh\sg}-\tfrac{1}{2}g_{\rh\sg}\pi\right)
   = m^2 \left( \pi_{\mu\nu} - g_{\mu\nu}\pi \right),
\end{multline}
where we have used the $\pi_{\mu\nu}$ equation of motion \eqref{C2pieom}
to simplify some terms. From the two equations of motion \eqref{C2pieom}
and \eqref{C2geom} it is clear that, although the action is not in
canonical form, we now have two propagating fields, the metric
$g_{\mu\nu}$ and the new auxiliary field $\pi_{\mu\nu}$. We expect the
particle content of the metric, which now satisfies a second-order
equation, to be the usual two helicity states. But does $\pi_{\mu\nu}$
really describe the degrees of freedom of a massive spin-two field? 

From our discussion in Section \ref{spin2}, we require generalized
divergence and trace conditions of the form \eqref{genconds} to hold
if this is the case. This is, in fact, so. Taking a trace of the
$g_{\mu\nu}$ field equation \eqref{C2geom} and the divergence of the
$\pi_{\mu\nu}$ equation of motion \eqref{C2pieom}, recalling that we
have the Bianchi identity $\nabla^\mu G_{\mu\nu}=0$, gives the pair of
conditions, 
\begin{equation}
\begin{aligned}
   \nabla^\mu \left( \pi_{\mu\nu} - g_{\mu\nu}\pi \right) &= 0, \\
   2\nabla^\mu\nabla^\nu \left( \pi_{\mu\nu} - g_{\mu\nu}\pi \right)
      &= -3 m^2 \pi. 
\end{aligned}
\end{equation}
These imply that 
\begin{equation}
\label{genpicond}
   \nabla^\mu \pi_{\mu\nu} = 0, \qquad 
   \pi = 0.
\end{equation}
These conditions have precisely the required form. In fact, they are
just the simplest curved-space generalization of the
free Pauli-Fierz conditions \eqref{spin2conds}. The ordinary
derivatives have simply been replaced by covariant derivatives. More
importantly they demonstrate that the full non-linear $R+C^2$ theory
is equivalent to a pure spin-two field coupled to gravity.

However, the action \eqref{C2auxS} is still not in a canonical
form. We would like both to grow an explicit kinetic term for the
auxiliary field and to reduce the curvature terms to the canonical
Einstein-Hilbert form. This should allow us to identify the mass of
the spin-two field, and to determine if it is indeed ghost-like. To do
this, we need to generalize the conformal transformation used in the
previous section for a scalar auxiliary field. We start by
rewriting the second-order action \eqref{C2auxS} in the suggestive
form, 
\begin{equation}
   S = \frac{1}{2\kappa^2} \int{ d^4x \sqrt{-g} \left[
        \left( \left[1+\tfrac{1}{2}\pi\right]g^{\mu\nu} - \pi^{\mu\nu}
           \right) R_{\mu\nu} 
        + \tfrac{1}{4}m^2 \left( \pi_{\mu\nu}\pi^{\mu\nu} - \pi^2 \right)
       \right] }. 
\end{equation}
It appears that to obtain a canonical Einstein-Hilbert term we need to define 
a new metric $\bg_{\mu\nu}$ such that 
\begin{equation}
   \sqrt{-\bg}\bg^{\mu\nu} = 
       \sqrt{-g} \left[ \left(1+\tfrac{1}{2}\pi\right)g^{\mu\nu} - \pi^{\mu\nu}
           \right].
\end{equation}
The necessary transformation can be written as
\begin{equation}
\begin{aligned}
\label{bgdef}
   \bg^{\mu\nu} &= \left(\textrm{det}A\right)^{-1/2} 
        g^{\mu\lm} A_\lm^{\ \ \nu}, \\
   A_\lm^{\ \ \nu} = A_\lm^{\ \ \nu} (\phi_{\sg\tau}) 
        &= \left(1+\tfrac{1}{2}\phi\right) \delta_\lm^{\ \ \nu}
             - \phi_\lm^{\ \ \nu}.
\end{aligned}
\end{equation}
Here we have introduced the new field
\begin{equation}
   \phi_\mu^{\ \ \nu} = \pi_\mu^{\ \ \nu},
\end{equation}
where it is understood that the indices of $\phi_{\mu\nu}$ are raised and 
lowered using the metric $\bg_{\mu\nu}$, while the indices of $\pi_{\mu\nu}$ 
where raised and lowered using $g_{\mu\nu}$. Although the identification is 
via the mixed index objects, it can be shown that $\phi_{\mu\nu}$ is 
nonetheless symmetric. Note also that if $\pi_{\mu\nu}$ is traceless, so is 
$\phi_{\mu\nu}$.

The transformation can be inverted to give
\begin{equation}
\label{invtransf}
   g_{\mu\nu} = g_{\mu\nu}(\phi_{\rh\sg})
        = \left(\det A\right)^{-1/2} A_\mu^{\ \ \lm} \bg_{\lm\nu}.
\end{equation}
so that the transformation of the Ricci tensor is given by
\begin{equation}
   R_{\mu\nu} = \bR_{\mu\nu} - \bnabla_\mu C^\lm_{\ \ \lm\nu}
         + \bnabla_\lm C^\lm_{\ \ \mu\nu}
         - C^\lm_{\ \ \mu\rh} C^\rh_{\ \ \nu\lm}
         + C^\lm_{\ \ \mu\nu} C^\rh_{\ \ \rh\lm},
\end{equation}
where $\bnabla_{\lm}\bg_{\mu\nu}=0$, 
\begin{equation}
   C^\lm_{\ \ \mu\nu} = C^\lm_{\ \ \mu\nu}(\phi_{\rh\sg})
       = \tfrac{1}{2}\left(g^{-1}\right)^{\lm\rh} \left( 
            \bnabla_\mu g_{\nu\rh} + \bnabla_\nu g_{\mu\rh}
            - \bnabla_\rh g_{\mu\nu} \right).
\end{equation}
and $g_{\mu\nu} = g_{\mu\nu}(\phi_{\rh\sg})$ as given in
\eqref{invtransf}. Thus in terms of the new variables
$(\bg_{\mu\nu},\phi_{\mu\nu})$, dropping a total divergence, the
action becomes 
\begin{multline}
\label{canRicci2}
   S = \frac{1}{2\kappa^2} \int{d^4x} \sqrt{-\bg} \left[ \bR
           - \bg^{\mu\nu} \left( C^\lm_{\ \ \mu\rh}(\phi_{\sg\tau}) 
                  C^\rh_{\ \ \nu\lm}(\phi_{\sg\tau})
              - C^\lm_{\ \ \mu\nu}(\phi_{\sg\tau}) 
                  C^\rh_{\ \ \rh\lm}(\phi_{\sg\tau}) \right)
              \right.  \\ \left. 
           + \tfrac{1}{4}m^2 \left(\textrm{det}A(\phi_{\sg\tau})\right)^{-1/2}
              \left( \phi_{\mu\nu}\phi^{\mu\nu} - \phi^2 \right)
           \right].
\end{multline}

\begin{figure}[ht]
   \centerline{\psfig{figure=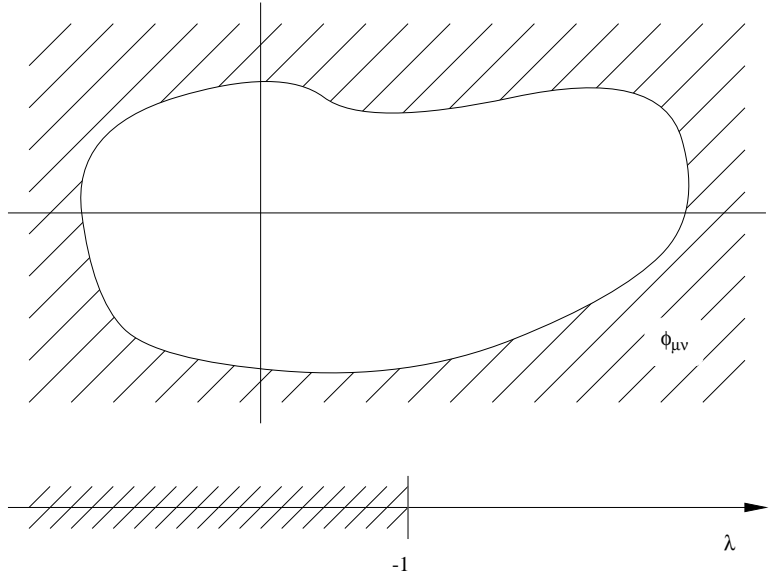,height=3in}}
   \caption{Schematic representation of the region in $\phi_{\mu\nu}$
       space giving conventional signature for the metric
       $\bg_{\mu\nu}$ compared with the same region for the field
       $\lm$} 
   \label{fig:phidomain}
\end{figure}

We recall that in the last section there was a subtlety in making the final 
transformation to the canonical form of the action. For the range of 
the auxiliary field $\lm<-1$, we had to define the field $\chi=\log(-1-\lm)$ 
which led to an action with the canonical form but the wrong overall 
normalization. Further at the special point $\lm=-1$ we could not transform 
the action into a canonical form. A similar subtlety appears here. From the 
form of the definition of $\bg_{\mu\nu}$ \eqref{bgdef}, we see that the 
new metric $\bg_{\mu\nu}$ need not have the same signature as the original 
metric $g_{\mu\nu}$. The definition also becomes undefined at the points in 
$\phi_{\mu\nu}$-space where $\det A=0$. We would like to be able to 
restrict ourselves to a range of $\phi_{\mu\nu}$ where the signature of the 
transformed metric is the conventional $(-,+,+,+)$, in agreement with the 
signature of the original metric. One way to do this is to note that, as we 
range in $\phi_{\mu\nu}$-space, a point where the $\bg_{\mu\nu}$ metric 
changes sign is indicated by $\det\bg_{\mu\nu}$ going to zero. From the 
definition \eqref{bgdef}, we see that
\begin{equation}
   \det \bg_{\mu\nu} = {\det A} \det g_{\mu\nu}.
\end{equation}
Thus since we assume $\det g_{\mu\nu}\neq0$, we find that the
signature-changing points are indicated by $\det A=0$. We also see
from the definition \eqref{bgdef} that when $\phi_{\mu\nu}=0$, we have
$\bg_{\mu\nu}=g_{\mu\nu}$ and so the signatures necessarily agree. 
These facts allow us to define a region of the $\phi_{\mu\nu}$-space
in which the $\bg_{\mu\nu}$ signature is conventional. It is the
region around the point $\phi_{\mu\nu}=0$ bounded by the surface $\det
A=0$. This region is shown schematically in
Figure~\ref{fig:phidomain}. The unshaded region where the signature is
conventional is the analog of the $\lm>-1$ region in the scalar field
case, the boundary $\det A=0$ is the analog of $\lm=-1$, while the
excluded region is the analog of $\lm<-1$. There may, of course, be
other disconnected regions where the signatures also agree, but here
we shall not consider them. It may also be possible to find a suitable
transformation in the regions where the signatures do not agree, just
as we defined $\chi=\log(-1-\lm)$ in the region $\lm<-1$ in the last
section. However, again as in the scalar field case, the resulting
action will not describe Einstein gravity; it will have either the
wrong normalization or a metric with the wrong signature. Here, and
throughout the paper, we will restrict our attention to that region of
the auxiliary field around $\phi_{\mu\nu}=0$ for which we know the
transformed action \eqref{canRicci2} describes conventional Einstein
gravity.

The equations of motion for the action \eqref{canRicci2} are extremely 
involved. However, the spin-two conditions on $\phi_{\mu\nu}$ can be 
obtained directly from equations \eqref{genpicond}. In terms of the new 
variables these become,
\begin{equation}
\begin{gathered}
   \bnabla^\mu\phi_{\mu\nu} - \bg^{\lm\mu}\left(
           C^\rh_{\ \ \lm\mu}(\phi_{\sg\tau}) \phi_{\rh\nu}
           + C^\rh_{\ \ \lm\nu}(\phi_{\sg\tau}) \phi_{\rh\mu} \right)
           = 0, \\
   \phi = 0,
\end{gathered}
\end{equation}
giving generalized divergence and trace conditions for $\phi_{\mu\nu}$
of the required form \eqref{genconds}, so that we may again conclude
that $\phi_{\mu\nu}$ describes only spin-two degrees of freedom. One
can also, of course, obtain these conditions directly from the
equations of motion of action \eqref{canRicci2}. 

Thus, we have succeeded in rewriting the original higher-derivative theory 
as a canonical theory of a spin-two field coupled to Einstein
gravity. The spin-two field has a generalized sigma-model kinetic
energy given by the non-linear $C^{\lm}_{\ \ \mu\nu}$ terms, and a
particular potential given by the term proportional to $m^2$. 

To make the structure of the sigma model explicit, we expand the kinetic 
energy and potential terms as a power series in $\phi_{\mu\nu}$ around 
$\phi_{\mu\nu}=0$. We find that
\begin{align}
   \mathcal{L}_{\phi_{\mu\nu}} = &- \tfrac{1}{4} \left(1 - \phi 
          + \ldots\right)
          \left[  \bnabla_\mu\phi\bnabla^\mu\phi 
          - \bnabla_\mu\phi_{\nu\rh}\bnabla^\mu\phi^{\nu\rh}
          + 2\bnabla^\mu\phi_{\mu\nu}\bnabla^\nu\phi
          - 2\bnabla_\mu\phi_{\nu\rh}\bnabla^\rh\phi^{\nu\mu}
          \right] \notag \\  
      & \hspace*{-0.1in} + \left[ \tfrac{1}{2} \phi^{\mu\nu} \left(
          \bnabla_\mu\phi\bnabla^\rh\phi_{\rh\nu}
          + \bnabla^\rh\phi\bnabla_\mu\phi_{\nu\rh}
          + \bnabla_\mu\phi_{\rh\sg}\bnabla_\nu\phi^{\rh\sg}
          - \bnabla_\mu\phi\bnabla_\nu\phi
          - 2\bnabla_\mu\phi_{\rh\sg} \bnabla^\sg
          \phi_\nu^{\ \ \rh} \right) + \ldots \right] \notag \\ 
      & + \tfrac{1}{4} m^2 \left( 1 + \phi + \ldots \right)
             \left( \phi_{\mu\nu}\phi^{\mu\nu} - \phi^2 \right). 
\label{phiexpand}
\end{align}
We see that, to lowest order, we have the curved-space version of the 
Pauli-Fierz action, with ordinary derivatives replaced with covariant 
derivatives. In the flat-space limit this reproduces Stelle's linearized 
result. As in that case, the kinetic term clearly has the wrong
sign, so that the spin-two field is ghost-like. For $m^2>0$ the ghost
has Planck-scale mass $m$, while for $m^2<0$ it is tachyonic. We note
that higher-order corrections in $\phi_{\mu\nu}$ modify both the
potential and kinetic energy terms. 

This result only applies, of course, in an expansion around $\phi_{\mu\nu}=0$. 
A relevant question is whether the sigma-model corrections could change the
spin-two field from ghost to non-ghost if we expand around a different 
vacuum, one where $\phi_{\mu\nu}$ takes on a non-zero expectation value. To 
answer this question we must first find the vacuum solutions of the reduced 
theory. In this paper, we shall only consider vacua in which the 
$\phi_{\mu\nu}$ field is covariantly constant. As stated above, the equations 
of motion for $\bg_{\mu\nu}$ and $\phi_{\mu\nu}$ are extremely involved, and, 
because of the tensor structure of $\phi_{\mu\nu}$, the covariantly
constant constraint is insufficient to render the vacuum equations
tractable. However, these equations simplify dramatically if we only
consider $\phi_{\mu\nu}$ of the form
$\phi_{\mu\nu}=\tfrac14\phi\bg_{\mu\nu}$. The $\bg_{\mu\nu}$ and
$\phi_{\mu\nu}$ equations of motion then become 
\begin{gather}
\label{canRicci2geom}
   \bR_{\mu\nu} - \tfrac12\bg_{\mu\nu}\bR = 
         \frac3{32} \left(1+\tfrac14\phi\right)^{-2} \left[
                \bnabla_\mu\phi\bnabla_\nu\phi
                -\tfrac12\bg_{\mu\nu}\left(\bnabla\phi\right)^2 \right]
         - \tfrac32 \bg_{\mu\nu} V(\phi), \\
\label{canRicci2chieom}
   \left(1+\tfrac14\phi\right)^{-2} 
      \left[\bnabla_\mu\bnabla_\nu\phi-\bg_{\mu\nu}\bnabla^2\phi\right]
   + \tfrac14\left(1+\tfrac14\phi\right)^{-3}
      \left[\bnabla_\mu\phi\bnabla_\nu\phi
         -\tfrac12\bg_{\mu\nu}\left(\bnabla\phi\right)^2 \right]
   = -12\bg_{\mu\nu}\frac{dV}{d\phi},
\end{gather}
respectively, where we have the potential function 
\begin{equation}
\label{canRicci2V}
    V(\phi) = \frac{m^2\phi^2}{16\left(1+\tfrac14\phi\right)^2}.
\end{equation}
The covariantly constant condition implies that 
\begin{equation}
   \bnabla_\mu\phi = \partial_\mu\phi = 0.
\end{equation}
It follows that $\phi$ is a constant and, from equation 
\eqref{canRicci2chieom}, that it must extremize the potential. The 
$\bg_{\mu\nu}$ equation of motion implies, as in equation \eqref{R2vac} 
that the vacuum is a space of constant curvature, a deSitter or
anti-deSitter solution with 
\begin{equation}
   \bR = 6 V(\phi). 
\end{equation}
Thus these vacua are spaces of constant curvature. The original metric, 
$g_{\mu\nu}$ is related to $\bg_{\mu\nu}$ through equation \eqref{invtransf}. 
For $\phi_{\mu\nu}=\tfrac14\phi\bg_{\mu\nu}$, it follows that 
\begin{equation}
\label{vacbgdef}
   g_{\mu\nu} = \left(1+\tfrac14\phi\right)^{-1} \bg_{\mu\nu},
\end{equation}
and, hence, since $\phi_{\mu\nu}$ is constant, that  
\begin{equation}
   R = \left(1+\tfrac14\phi\right) \bR.
\end{equation}
Thus these vacua correspond to spaces of constant curvature with respect to 
the original metric as well, although the radius of curvature is in general 
different. We should note that equation \eqref{vacbgdef} provides us with a 
specific example of the problem of signature-changing transformations. 
When $\phi<-4$ the signature of $\bg_{\mu\nu}$ changes sign with respect to 
the signature of $g_{\mu\nu}$. This means we are moving outside the 
conventional-signature region around $\phi_{\mu\nu}=0$ we defined
above. Note that the boundary $\phi=-4$, where the transformation
becomes undefined, does indeed correspond to $\det A=0$. Thus
throughout this discussion of vacua of the form
$\phi_{\mu\nu}=\tfrac14\phi\bg_{\mu\nu}$ we must assume $\phi>-4$. 

\begin{figure}[ht]
   \centerline{\psfig{figure=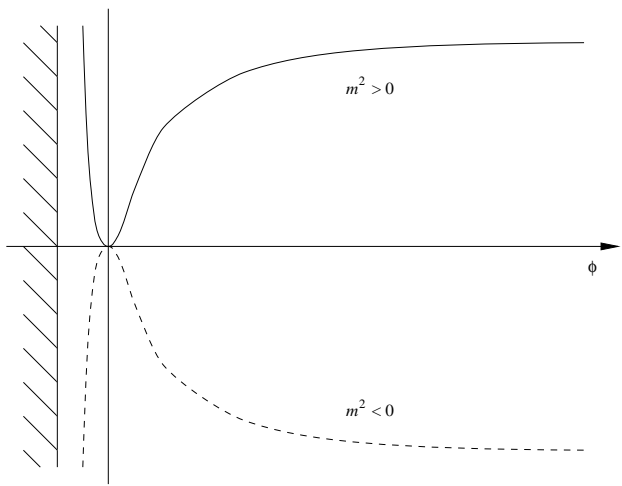,height=2.5in}}
   \caption{$V(\phi)$ for quadratic higher-derivative gravity}
   \label{fig:Ricci2V}
\end{figure}

The potential $V(\phi)$ is plotted in Figure \ref{fig:Ricci2V}. We see
that the only stationary point is at $\phi=0$, so that
$\phi_{\mu\nu}=0$. This is in fact simply a confirmation of our
general result that the equations of motion imply $\phi=0$. For
$m^2>0$ this is a stable minimum, whereas for $m^2<0$ it is
unstable. Since $V(0)=0$, it follows that $\bR=R=0$ and, hence, that
$\bg_{\mu\nu}=g_{\mu\nu}=\eta_{\mu\nu}$. We are led to conclude that
trivial flat space with $\phi_{\mu\nu}=0$ is the only possible stable
vacuum state of the type we are considering. Therefore the spin-two
excitations around the vacuum are necessarily ghost-like. 

In this section we have made the five new spin-two degrees of freedom
in $R+C^2$ gravity explicit by rewriting the theory in a truly
canonical second-order form. It is worth stressing that these degrees
of freedom are physical and cannot be removed by a field redefinition
or a gauge transformation. We have shown that the new field
$\phi_{\mu\nu}$ satisfies generalized spin-two conditions of the
required form. The second-order theory has an interesting and
complicated structure with a sigma-model kinetic energy and a fixed
potential. We have shown that the only symmetric vacuum state 
where $\phi_{\mu\nu}$ is covariantly constant
is flat space with $\phi_{\mu\nu}=0$. The fluctuations of
$\phi_{\mu\nu}$ around this vacuum, always have the wrong-sign kinetic
energy and so the ghost problem persists in the full non-linear
theory. 

\section{A Massless Spin-Two Theory Consistently Coupled to Gravity}
\label{m2=0}

We showed in the previous section that the $R+C^2$ action is equivalent
to an action describing a massive spin-two field coupled to
gravity. Given the problem of writing a consistent theory of a
spin-two field coupled to gravity, it is natural to ask if the
second-order theory has a sensible massless limit. Consider the
non-canonical form given in the last line of \eqref{C2auxS}. Setting
$m=0$, we have the action 
\begin{equation}
\label{mzeroS}
   S = \frac{1}{2\kappa^2} \int{ d^4x \sqrt{-g} \left[
       R - G_{\mu\nu}\pi^{\mu\nu} \right] }, 
\end{equation}
From the discussion in Section \ref{spin2}, if this is to describe a
consistent theory of spin-two, there must be a local symmetry of the
form \eqref{gensymm}. In fact, a very simple symmetry does exist. 
Consider the transformation 
\begin{equation}
\label{mzerosymm}
   \pi_{\mu\nu} \rightarrow \pi_{\mu\nu} + \nabla_{(\mu}\xi_{\nu)}. 
\end{equation}
Because of the Bianchi identity $\nabla^\mu G_{\mu\nu}=0$, integrating
by parts, we find that this is a symmetry of the action
\eqref{mzeroS}. Further it has the required form \eqref{gensymm}. It
is just the simple curved-space generalization of the free Pauli-Fierz
symmetry, with ordinary derivatives replaced by covariant
derivatives. We conclude that we have a consistent theory of a
massless spin-two field coupled to gravity. 

To understand how this consistency works, consider the equations of
motion following from \eqref{mzeroS}. The $\pi_{\mu\nu}$ field
equations are  
\begin{equation}
\label{mzerogeom}
   G_{\mu\nu} = 0, 
\end{equation}
which are just the Einstein equations for empty space, implying that
$R_{\mu\nu}=0$. The $g_{\mu\nu}$ field equations then reads 
\begin{equation}
\label{mzeropieom}
   \nabla^2\pi_{\mu\nu} + g_{\mu\nu}\nabla^\rh\nabla^\sg\pi_{\rh\sg}
      - \nabla_\mu\nabla^\lm\pi_{\lm\nu} - \nabla_\nu\nabla^\lm\pi_{\lm\mu}
      + \nabla_\mu\nabla_\nu\pi - g_{\mu\nu}\nabla^2\pi 
   = 0,
\end{equation}
where we have substituted by $R_{\mu\nu}$ throughout. But these are
simply the equations one would derive for the fluctuations of
the metric about a fixed empty-space background. That is, for a
background spacetime given by $g_{\mu\nu}$ satisfying
\eqref{mzerogeom}, linear perturbations of the metric satisfy the 
equations \eqref{mzeropieom}. Such fluctuations are known to be
consistent; they represent the graviton degrees of freedom in empty
space. This is the origin of the consistency of $\pi_{\mu\nu}$. Note,
however, the form of the action \eqref{mzeroS} implies
that unlike the metric perturbation the $\pi_{\mu\nu}$ field is 
ghost-like. 

It is worth noting that since gravity is also a consistent theory of a
massless spin-two field, the action \eqref{mzeroS} must have two local
symmetries. Indeed it does. One is the $\pi_{\mu\nu}$ symmetry given
above. The other is, of course, diffeomorphism invariance. We should
also be able to make contact with Wald's work on consistent
theories of collections of spin-two fields \cite{CQG-4-1279}. Wald identified
consistent theories with geometrical theories of a metric on an
algebra-valued manifold, where the algebra must be commutative and
associative. For an algebra with $n$ elements, the metric on the
algebra-valued manifold describes $n$ different spin-two fields. In
the simple case of an algebra consisting of only the identity
element $e$ and a single nil-potent element $v$, Wald showed that the
generalized algebra-valued Einstein-Hilbert action is really a pair of
actions. Expanding in the basis $(e,v)$, we have 
\begin{equation}
\begin{aligned}
   S_e &= \frac{1}{2\kappa^2} \int{ d^4x \sqrt{-g} R }, \\
   S_v &= \frac{1}{2\kappa^2} \int{ d^4x \sqrt{-g} G_{\mu\nu}\pi^{\mu\nu} },
\end{aligned}
\end{equation}
where $g_{\mu\nu}$ and $\pi_{\mu\nu}$ are the two components of the
algebra-valued metric. Thus we see that the action \eqref{mzeroS} is
the sum of the two actions of Wald's theory. Note that the two actions
really give three equations of motion for two fields, since we can
vary $g_{\mu\nu}$ in $S_e$ and $g_{\mu\nu}$ and $\pi_{\mu\nu}$ in
$S_v$. However, the $\pi_{\mu\nu}$ equation and the $g_{\mu\nu}$
equation from $S_e$, both give the empty space Einstein equations
$G_{\mu\nu}=0$. Thus the system is not over-constrained. Including the
$g_{\mu\nu}$ equation from $S_v$, exactly the same equations of
motion arise from the single action \eqref{mzeroS} as from the pair of
Wald actions. 

Finally, we point out that the massless theory can still be rewritten
in the canonical form \eqref{canRicci2}. The local symmetry
\eqref{mzerosymm} is of course preserved. However, it now has a more
complicated realization, involving a simultaneous transformation of
the metric and the spin-two field,
\begin{equation}
\begin{aligned}
   \phi_{\mu\nu} &\rightarrow 
        {\phi'}_{\mu\nu} = \phi_{\mu\nu} + \nabla_{(\mu}\xi_{\nu)} 
            - C^\rho_{\ \ \mu\nu}(\phi_{\sg\tau}) \xi_\rho, \\
   g_{\mu\nu} &\rightarrow 
        {g'}_{\mu\nu} = \left(1+\tfrac12\phi'\right)^{-1} \left[
                \frac{\det^{1/2} A(\phi'_{\sg\tau})}
                   {\det^{1/2} A(\phi_{\sg\tau})}
                \left(1+\tfrac12\phi\right)g_{\mu\nu} 
                + \left(\phi'_{\mu\nu}-\phi_{\mu\nu}\right)
             \right].
\end{aligned}
\end{equation}
Nonetheless, it is still in the required form \eqref{gensymm}. In
particular, it collapses to the old massless free Pauli-Fierz symmetry
in the linearized limit. In the $\phi_{\mu\nu}$-formulation, it is
clear that the problem remains that the spin-two field is
ghost-like. Wald has shown that all the algebra-valued theories he
derived include at least one ghost-field. Thus, although quadratic
gravity provides an interesting example of a massless spin-two field
consistently coupled to gravity, it does so at the expense of an
apparent loss of unitarity. 

\section{The General Quadratic Theory}
\label{gen2}

Having dealt separately with the two special cases of pure scalar-curvature  
and pure Weyl-tensor corrections to gravity, we now turn to 
extracting the new degrees of freedom for a general quadratic action. The 
goal is to separate both a new spin-two and a new scalar field 
from the full non-linear action. As we have already discussed, using the 
Gauss-Bonnet theorem, the general action can be written either in terms of 
scalar-curvature and Weyl-tensor terms, or equivalently in terms of 
scalar-curvature and Ricci-tensor terms. We have
\begin{align}
   S &= \frac{1}{2\kappa^2} \int{ d^4x \sqrt{-g} 
        \left[ R + \frac{1}{6{m_0}^2} R^2 - \frac{1}{2{m_2}^2} 
            C^2 \right] } \notag \\
     &= \frac{1}{2\kappa^2} \int{ d^4x \sqrt{-g} 
        \left[ R + \frac{1}{6{m_0}^2} R^2 - \frac{1}{{m_2}^2} 
            \left( R_{\mu\nu}R^{\mu\nu} - \tfrac{1}{3}R^2 \right)
        \right] }. 
\end{align}
Having seen how to introduce auxiliary fields for these two types of
term in the last two subsections, it is very natural simply to repeat
the procedure here with a pair of auxiliary fields, a scalar for the
scalar-curvature correction term and a tensor for the Weyl term. Thus
combining the rewritings \eqref{R2auxS} and 
\eqref{C2auxS} we have
\begin{align}
   S &= \frac{1}{2\kappa^2} \int{ d^4x \sqrt{-g} 
        \left[ R + \frac{1}{6{m_0}^2} R^2 - \frac{1}{{m_2}^2} 
            \left( R_{\mu\nu}R^{\mu\nu} - \tfrac{1}{3}R^2 \right)
        \right] } \notag \\
     &= \frac{1}{2\kappa^2} \int{ d^4x \sqrt{-g} \left[
           R + \lm R - \tfrac32 {m_0}^2 \lm^2
           - G_{\mu\nu}\pi^{\mu\nu} + \tfrac{1}{4}{m_2}^2
              \left(\pi_{\mu\nu}\pi^{\mu\nu} - \pi^2 \right)
         \right] } \notag \\
     &= \frac{1}{2\kappa^2} \int{ d^4x \sqrt{-g} \left[
         e^\chi R - \tfrac{3}{2}{m_0}^2\left(e^\chi-1\right)^2
         - G_{\mu\nu}\pi^{\mu\nu} + \tfrac{1}{4}{m_2}^2
            \left(\pi_{\mu\nu}\pi^{\mu\nu} - \pi^2 \right)
         \right] },
\label{fullauxS}
\end{align}
where in the last line, as before, we have defined a new variable 
$\chi=\log\left(1+\lm\right)$. For the time being we have not made 
the usual accompanying conformal rescaling of the metric. The equations of 
motion for $\chi$ and $\pi_{\mu\nu}$ are, exactly as before,
\begin{equation}
\begin{aligned}
   R &= 3 {m_0}^2 \left( e^\chi - 1 \right), \\
   G_{\mu\nu} &= \tfrac12 {m_2}^2 \left(\pi_{\mu\nu}-g_{\mu\nu}\pi\right), 
\end{aligned}
\end{equation}
so that substituting by the $\pi_{\mu\nu}$ equation reproduces the Weyl-tensor 
term, while substituting by the $\chi$ equation reproduces the squared 
curvature-scalar term.

The $g_{\mu\nu}$ equation of motion is 
\begin{multline}
   2e^\chi G_{\mu\nu} - 2\left(\nabla_\mu\nabla_\nu-g_{\mu\nu}\nabla^2\right)
             e^\chi + \tfrac32 g_{\mu\nu}{m_0}^2\left(e^\chi-1\right)^2 \\
      + \nabla^2\pi_{\mu\nu} + g_{\mu\nu}\nabla^\rh\nabla^\sg\pi_{\rh\sg}
      - \nabla_\mu\nabla^\lm\pi_{\lm\nu} - \nabla_\nu\nabla^\lm\pi_{\lm\mu}
      + \nabla_\mu\nabla_\nu\pi - g_{\mu\nu}\nabla^2\pi \\
      + R_\mu^{\ \ \lm} \left(\pi_{\lm\nu}-\tfrac{1}{2}g_{\lm\nu}\pi\right)
      + R_\nu^{\ \ \lm} \left(\pi_{\lm\mu}-\tfrac{1}{2}g_{\lm\mu}\pi\right)
      - \tfrac{1}{2}g_{\mu\nu}R^{\rh\sg}
             \left(\pi_{\rh\sg}-\tfrac{1}{2}g_{\rh\sg}\pi\right) = 0.
\end{multline}
Taking a trace of this equation and a divergence of the $\pi_{\mu\nu}$ 
equation, after using the $\chi$ equation to remove all $R$ dependence,
gives the conditions
\begin{equation}
\label{niaveconds}
\begin{aligned}
   \nabla^\mu \left( \pi_{\mu\nu} - g_{\mu\nu}\pi \right) &= 0, \\
   \pi - 2{m_2}^2 \nabla^2 e^\chi &= 0. 
\end{aligned}
\end{equation}
It appears that we have derived a set of spin-two conditions for
$\pi_{\mu\nu}$. However, they do not have the correct linearized
form. Rather than collapsing to the free conditions
$\partial^\mu\pi_{\mu\nu}=\pi=0$, we find there is a linear dependence
on the scalar field $\chi$. 

This linear dependence is a consequence of the fact that it is not
possible, when the auxiliary fields are introduced in this way, to
separate the scalar and spin-two kinetic energy terms at the quadratic
level. To make this coupling explicit, consider making the usual
conformal transformation, $\tg_{\mu\nu}=e^\chi g_{\mu\nu}$, to grow a
kinetic term for the scalar field. The Einstein tensor is not
invariant but grows terms depending on the derivative of $\chi$,
giving 
\begin{multline}
   S = \frac{1}{2\kappa^2} \int{ d^4x \sqrt{-\tg}} \left[ \tR 
         - \tfrac{3}{2}\left(\tnabla\chi\right)^2
         - \tfrac{3}{2}{m_0}^2\left(1-e^{-\chi}\right)^2
         - \tG_{\mu\nu}\pi^{\mu\nu} 
         \right. \\ \left.
         - \left( \tnabla_\mu\tnabla_\nu\chi - \tg_{\mu\nu}\tnabla^2\chi
            + \tnabla_\mu\chi\tnabla_\nu\chi 
            - \tg_{\mu\nu}\left(\tnabla\chi\right)^2 \right) \pi^{\mu\nu}
         + \tfrac{1}{4}{m_2}^2
            \left(\pi_{\mu\nu}\pi^{\mu\nu} - \pi^2 \right)
         \right],
\end{multline}
where now the indices of $\pi_{\mu\nu}$ are raised and lowered with the 
metric $\tg_{\mu\nu}$, and we take the object with both indices lowered as 
invariant when we change metrics. We see that because of the transformation 
of the Einstein tensor, there are couplings between 
$\tnabla_\mu\tnabla_\nu\chi$ and $\pi_{\mu\nu}$
at the quadratic level. Clearly if we now try and grow kinetic 
terms for the spin-two field by moving to a new metric $\bg_{\mu\nu}$, 
just as in the case of pure Weyl-squared corrections, this quadratic coupling 
between the kinetic energy terms for $\pi_{\mu\nu}$ and $\chi$ will persist. 
Further this coupling cannot be removed by simple field redefinition. Thus 
the spin-two part of this action cannot reduce to the Pauli-Fierz action 
in the quadratic limit. 

Although there is nothing wrong with the previous formulation, we would 
prefer to have a more canonical form giving the ordinary Pauli-Fierz 
limit. We notice that the quadratic kinetic coupling arose because of the 
inhomogeneous pieces in the transformation of the Einstein tensor under 
a conformal rescaling. One way to circumvent such inhomogeneous terms is to 
recall that the Weyl-squared term is invariant under conformal 
rescalings. Thus an alternative procedure is to introduce only the scalar 
auxiliary field, leaving the Weyl-squared terms untouched, make the conformal 
rescaling to grow kinetic terms for the scalar field, and only then to 
introduce the spin-two auxiliary field. That is,
\begin{equation}
\begin{split}
\label{twostepauxS}
   S &= \frac{1}{2\kappa^2} \int{ d^4x \sqrt{-g} \left[ 
           R + \frac{1}{6{m_0}^2} R^2 
           - \frac{1}{2{m_2}^2} C_{\lm\mu\nu\rh}C^{\lm\mu\nu\rh} 
           \right] } \\
     &= \frac{1}{2\kappa^2} \int{ d^4x \sqrt{-g} \left[ e^\chi R 
           - \tfrac32 {m_0}^2\left(e^\chi-1\right)^2
           - \frac{1}{2{m_2}^2} C_{\lm\mu\nu\rh}C^{\lm\mu\nu\rh} 
           \right] } \\
     &= \frac{1}{2\kappa^2} \int{ d^4x \sqrt{-\tg} \left[ \tR 
           - \tfrac32 \left(\tnabla\chi\right)^2 
           - \tfrac32 {m_0}^2\left(1-e^{-\chi}\right)^2
           - \frac{1}{2{m_2}^2} \tC_{\lm\mu\nu\rh}\tC^{\lm\mu\nu\rh}
           \right] } \\
     &= \frac{1}{2\kappa^2} \int{ d^4x \sqrt{-\tg} \left[ \tR 
           - \tfrac32 \left(\tnabla\chi\right)^2 
           - \tfrac32 {m_0}^2\left(1-e^{-\chi}\right)^2
           - \tG_{\mu\nu}\tpi^{\mu\nu} 
           + \tfrac14 {m_2}^2 \left( \tpi_{\mu\nu}\tpi^{\mu\nu} - \tpi^2 
\right)
           \right] },
\end{split}
\end{equation}
where, in the third line, we have made a conformal transformation to
the metric $\tg_{\mu\nu}=e^\chi g_{\mu\nu}$, and as usual dropped a
Gauss-Bonnet term to rewrite the Weyl-squared term in the third line
as $R_{\mu\nu}R^{\mu\nu}-\tfrac13R^2$ before introducing the auxiliary
field $\tpi_{\mu\nu}$.  Clearly there is now no quadratic coupling
between $\chi$ and $\pi_{\mu\nu}$. We can again derive, from the trace
of the $\tg_{\mu\nu}$ equations of motion and the divergence of the
$\tpi_{\mu\nu}$ equations of motion, the generalized divergence and
trace conditions on $\tpi_{\mu\nu}$. They are given by  
\begin{equation}
\label{tpicond}
\begin{gathered}
   \tnabla^{\mu} \left( \tpi_{\mu\nu} - g_{\mu\nu}\tpi \right) = 0, \\
   \tpi - {m_2}^{-2} \left[ \left(\tnabla\chi\right)^2 
            + 2{m_0}^2 \left(1-e^{-\chi}\right)^2 \right] = 0.
\end{gathered}
\end{equation}
Linearizing, we now find that the conditions have no dependence
on the scalar field, collapsing to the free Pauli-Fierz
conditions. We can conclude that they provide generalized spin-two
conditions of the correct form \eqref{genconds}, and that this
parametrization provides a true separation of the scalar and spin-two
degree of freedom. 

To complete the transformation to canonical form, we define a new metric, 
exactly as in \eqref{bgdef}, but with $g_{\mu\nu}$ replaced with 
$\tg_{\mu\nu}$, and obtain the action
\begin{multline}
\label{canongen}
   S = \frac{1}{2\kappa^2} \int{d^4x} \sqrt{-\bg} \left[ \bR
           - \tfrac{3}{2}\left(A^{-1}(\phi_{\sg\tau})\right)_{\mu}^{\ \ \nu}
                  \bnabla^\mu\chi\bnabla_\nu\chi
           - \tfrac{3}{2}\left(\textrm{det}A(\phi_{\sg\tau})\right)^{-1/2}
                  \left(1-e^{-\chi}\right)^2
              \right. \\ \left. 
           - \bg^{\mu\nu} \left( C^\lm_{\ \ \mu\rh}(\phi_{\sg\tau}) 
                  C^\rh_{\ \ \nu\lm}(\phi_{\sg\tau})
              - C^\lm_{\ \ \mu\nu}(\phi_{\sg\tau}) 
                  C^\rh_{\ \ \rh\lm}(\phi_{\sg\tau}) \right)
              \right. \\ \left. 
           + \tfrac{1}{4}{m_2}^2 
              \left(\textrm{det}A(\phi_{\sg\tau})\right)^{-1/2}
              \left( \phi_{\mu\nu}\phi^{\mu\nu} - \phi^2 \right)
           \right].
\end{multline}

We have a canonical form for a spin-two field and a scalar field 
coupled to gravity. The spin-two field now couples to the scalar kinetic 
energy. It also enters the scalar potential. In both cases, however, the 
coupling is at the cubic level, so that the linearized field equations 
decouple, and, exactly as for the pure Weyl-squared case
\eqref{phiexpand} expanding to quadratic order about
$\phi_{\mu\nu}=0$ in flat space, gives the Pauli-Fierz
action. Furthermore, the spin-two conditions \eqref{tpicond}, in the
new $\bg_{\mu\nu}$ frame become 
\begin{equation}
\begin{gathered}
   \bnabla^\mu \left(\phi_{\mu\nu}-\bg_{\mu\nu}\phi\right) 
       - \bg^{\lm\mu}C^\rh_{\ \ \lm\mu}(\phi_{\sg\tau}) 
             \left(\phi_{\rh\nu}-\bg_{\rh\nu}\phi\right)
       - \bg^{\lm\mu}C^\rh_{\ \ \lm\nu}(\phi_{\sg\tau}) 
             \left(\phi_{\rh\mu}-\bg_{\rh\mu}\phi\right)
        = 0, \\
   \phi - {m_2}^{-2} \left[ \left(\det A(\phi_{\sg\tau})\right)^{1/2} 
        \left(A^{-1}(\phi_{\sg\tau})\right)_{\mu}^{\ \ \nu}
             \bnabla^\mu\chi\bnabla_\nu\chi
        + 2{m_0}^2 \left(1-e^{-\chi}\right)^2 \right] = 0,
\end{gathered}
\end{equation}
and, although complicated, are still of the correct form
\eqref{genconds} to imply that $\phi_{\mu\nu}$ is pure spin-two. 

We can again ask about the vacuum state of the theory. As before we look 
for stable solutions of the equations of motion, where the auxiliary 
fields are covariantly constant. To render the problem tractable we 
again impose the additional condition that 
$\phi_{\mu\nu}=\tfrac14\phi\bg_{\mu\nu}$. The equations of motion then 
greatly simplify to give
\begin{multline}
   \bR_{\mu\nu} - \tfrac12\bg_{\mu\nu}\bR = 
       \tfrac32\left(1+\tfrac14\phi\right)^{-1}\left[
           \bnabla_\mu\chi\bnabla_\nu\chi
           -\tfrac12\bg_{\mu\nu}\left(\bnabla\chi\right)^2 \right] \\
       + \frac3{32} \left(1+\tfrac14\phi\right)^{-2} \left[
           \bnabla_\mu\phi\bnabla_\nu\phi
           -\tfrac12\bg_{\mu\nu}\left(\bnabla\phi\right)^2 \right]
       - \tfrac32 \bg_{\mu\nu} V(\chi,\phi), 
\end{multline}
\begin{equation}
   \left(1+\tfrac14\phi\right)^{-1}\bnabla^2\chi
       - \tfrac14\left(1+\tfrac14\phi\right)^{-2}
           \bnabla^\mu\phi\bnabla_\mu\chi 
       = \frac{dV}{d\chi}, 
\end{equation}
\begin{multline}
   \left(1+\tfrac14\phi\right)^{-2} 
           \left[\bnabla_\mu\bnabla_\nu\phi-\bg_{\mu\nu}\bnabla^2\phi\right]
       + \tfrac14\left(1+\tfrac14\phi\right)^{-3}
           \left[\bnabla_\mu\phi\bnabla_\nu\phi
              -\tfrac12\bg_{\mu\nu}\left(\bnabla\phi\right)^2 \right] \\
       = 6 \left[ \bnabla_\mu\chi\bnabla_\nu\chi
              -\tfrac12\bg_{\mu\nu}\left(\bnabla\chi\right)^2 \right]
       - 12\bg_{\mu\nu}\frac{dV}{d\phi},
\end{multline}
respectively, where we have introduced the potential 
\begin{equation}
    V(\chi,\phi) = \frac{{m_0}^2\left(1-e^{-\chi}\right)^2}
              {2\left(1+\tfrac14\phi\right)^2}
         + \frac{{m_2}^2\phi^2}{16\left(1+\tfrac14\phi\right)^2}.
\end{equation}
The covariantly constant conditions imply that  
\begin{equation}
\begin{aligned}
   \bnabla_\mu\chi &= \partial_\mu\chi = 0, \\
   \bnabla_\mu\phi &= \partial_\mu\phi = 0,
\end{aligned}
\end{equation}
and from the equations of motion we see that the vacuum must be an 
extremum of the potential. It is easy to show that the only solution is 
again trivial flat-space with $\chi=\phi_{\mu\nu}=0$. Following exactly the 
discussion for the pure Weyl-squared theory, we find that the spin-two 
excitations around this vacuum are ghost-like, while it is clear from the 
form of the action \eqref{canongen} that the graviton and scalar excitations 
are normal. Thus, the problem of the ghost-like spin-two field persists even 
for general quadratic actions. 

Finally we would like to understand what gave us the freedom to introduce two 
different auxiliary spin-two fields, $\tpi_{\mu\nu}$ which gave the canonical 
Pauli-Fierz quadratic limit and $\pi_{\mu\nu}$ which did not. The two 
fields are not related by a simple field redefinition, but rather, given 
their respective trace conditions, appear to differ by terms which are 
derivatives of $\chi$. The solution is that there is an extra gauge freedom 
present when introducing the spin-two auxiliary field, the
generalization of the St\"uckleberg symmetry discussed in Section
\ref{spin2}. Specifically, instead of the last line of
\eqref{fullauxS} we can introduce a new field $\zeta_\mu$ and write,
\begin{multline}
\label{stuckS}
   S = \frac{1}{2\kappa^2} \int{ d^4x \sqrt{-g} \left[ e^\chi R + 
           - \tfrac{3}{2}{m_0}^2\left(e^\chi-1\right)^2 
           - G_{\mu\nu}\pi^{\mu\nu} 
           \right. } \\ \left. 
           + \tfrac{1}{4}m^2 \left( 
              \left[\pi_{\mu\nu}+\nabla_{(\mu}\zeta_{\nu)}\right]
              \left[\pi^{\mu\nu}+\nabla^{(\mu}\zeta^{\nu)}\right]
              - \left[\pi+\nabla^\mu\zeta_\mu\right]^2 \right)
           \right].
\end{multline}
The presence of the field $\zeta_\mu$ has no effect on eliminating 
$\pi_{\mu\nu}$ through its own equation of motion, since using the Bianchi 
identity $\nabla^\mu G_{\mu\nu}=0$, its only contribution, on substituting for 
$\pi_{\mu\nu}$, is a total divergence. Further, 
its field equation, $\nabla^\mu\left(\pi_{\mu\nu}-g_{\mu\nu}\pi\right)=0$, is 
already implied by the $\pi_{\mu\nu}$ equation of motion. The action
\eqref{stuckS} is thus completely equivalent to the action in the last
line of \eqref{fullauxS}. More significantly, by including the field
$\zeta_\mu$, we introduce a new gauge symmetry. Namely the action is
invariant under the combined transformations 
\begin{equation}
\label{genStucksymm}
\begin{aligned}
   \pi_{\mu\nu} &\rightarrow \pi_{\mu\nu} + \nabla_{(\mu}\xi_{\nu)}, \\
   \zeta_\mu &\rightarrow \zeta_\mu - \xi_\mu.
\end{aligned}
\end{equation}
In this language, the original expression \eqref{fullauxS}
corresponds to the gauge choice $\zeta_\mu=0$. As in the free case, the
divergence condition \eqref{niaveconds} becomes the conserved current
condition (in the $\zeta_\mu=0$ gauge) for the symmetry
\eqref{genStucksymm}. 

We can now understand the connection between the two spin-two fields 
$\pi_{\mu\nu}$ and $\tpi_{\mu\nu}$. They are related by gauge transformations, 
though in a rather subtle way since they are defined with respect to two 
different metrics. $\pi_{\mu\nu}$ is defined in the $g_{\mu\nu}$ frame while 
$\tpi_{\mu\nu}$ is defined in the conformally rescaled $\tg_{\mu\nu}$ frame. 
To see the relationship explicitly, we start by choosing the gauge 
$\zeta_\mu={m_2}^{-2}\nabla_\mu\chi$ in the action 
\eqref{stuckS}. As already stated, this is the same action as given in 
\eqref{fullauxS}, just expressed in a different gauge. If we now make the 
conformal rescaling to $\tg_{\mu\nu}=e^\chi g_{\mu\nu}$, as before, the 
$G_{\mu\nu}\pi^{\mu\nu}$ term generates derivative terms coupling 
$\pi_{\mu\nu}$ with $\chi$. However, with this particular gauge choice, we 
find that such terms so arrange themselves to give, after some algebra and 
dropping a total derivative term,
\begin{multline}
   S = \frac{1}{2\kappa^2} \int{ d^4x \sqrt{-\tg} \left[ \tR 
           - \tfrac{3}{2}\left(\tnabla\chi\right)^2
           - \tfrac{3}{2}{m_0}^2\left(1-e^{-\chi}\right)^2
           - \tG_{\mu\nu}\pi^{\mu\nu} 
           \right. } \\ \left. 
           + \tfrac{1}{4}m^2 \left( 
              \left[\pi_{\mu\nu}-{m_2}^{-2}\tnabla_\mu\tnabla_\nu\chi\right]
              \left[\pi^{\mu\nu}-{m_2}^{-2}\tnabla^\mu\tnabla^\nu\chi\right]
              - \left[\pi-{m_2}^{-2}\tnabla^2\chi\right]^2 \right)
           \right].
\end{multline}
We recognize this as none other than the action in the last line of 
\eqref{twostepauxS}, but instead of the gauge $\tzeta_\mu=0$, we have 
$\tzeta_\mu=-{m_2}^{-2}\tnabla_\mu\chi$. The subtlety is that the new action 
is gauge-fixed with respect to the new metric $\tg_{\mu\nu}$, that is, we 
have $\tnabla_\mu\tzeta_\nu$ not $\nabla_\mu\tzeta_\nu$ entering the action. 
Thus, the procedure of introducing the auxiliary fields in two stages using 
the conformal invariance of the Weyl terms, is equivalent to first choosing 
a particular gauge in the $g_{\mu\nu}$ frame. Then, with this choice, the 
transformed action arranges into a (different) gauge-fixed form in the new 
$\tg_{\mu\nu}$ frame, thereby removing the coupling between the spin-two 
and scalar kinetic terms. This decomposition is not unique, but has advantage 
of also largely decoupling the kinetic energies for the two fields in the 
final nonlinear expression.

In summary, in this section we have been able to rewrite the general action 
quadratic in the curvature tensors in a canonical form \eqref{canongen}, which
makes the new scalar and spin-two degrees explicit at the non-linear
level. The action has the correct quadratic limit, and we have shown
that the spin-two field satisfies a pair of generalized conditions
which constrain the number of independent components to the five
degrees of freedom of a massive spin-two field. In the linearized
limit these conditions become the trace and divergence conditions of
the free Pauli-Fierz equation of motion. We found that the theory has
a single vacuum state, when the fields are covariantly constant, 
at $\chi=\phi_{\mu\nu}=0$, with the mass of the
scalar and spin-two excitations generically of the order of the
Planck-scale. Further, we made explicit the ghost nature of the
spin-two field about this vacuum, demonstrating that this problem
persists in the full non-linear theory. 

\section{Conclusion}
\label{concl}

As we discussed in the introduction, this paper has centered on two
questions. One is how to rewrite quadratic gravity in a canonical
second-order form, extracting the new scalar and spin-two degrees of
freedom. The other is how such a canonical theory provides a
consistent description of a coupled spin-two field. Two interesting
questions arise. The first is to understand the connection between
Wald's geometrical theories of multiple spin-two fields and
higher-derivative gravitation. In this context, one might consider
actions which involve higher derivatives of the curvature
tensors. This would imply more degrees of freedom, either new scalar
or spin-two fields. If such theories can be rewritten in a canonical
second-order form, they should provide examples of consistently
coupled theories with multiple spin-two fields. It would be
interesting to see how Wald's algebras are realized in such theories,
and what additional structure they imply for the higher-derivative
theory.

The second question is whether non-trivial
vacua exist in more general theories of higher-derivative
gravity. Naively, all such theories represent Planck-scale corrections
to Einstein gravity and so apparently have no effect on low-energy
physics. However, this assumes that flat space is the only vacuum
state in the theory. Generally one must always use the full theory to
find the vacuum states and only then, having chosen a particular
vacuum, make the low-energy expansion. Thus
higher-derivative terms cannot be neglected when determining the
correct vacuum state for the low-energy physics. For quadratic
theories we found that the only stable vacuum was trivial flat
space. To investigate more complicated theories, we can essentially
use the same techniques we have developed in this paper. (Some
discussion of reducing more general higher-derivative theories of
gravity to a second-order form has already been given in the
literature \cite{GRG-19-465,CQG-5-L95,CQG-7-557,PRD-37-1406,PLB-214-515,%
CQG-7-893,CQG-7-1023,CQG-11-269}.) Recently, this has allowed us to show that
non-trivial vacua seem to be a generic feature of more general
theories \cite{PRD-53-5597}, indicating that higher-derivative
corrections to gravity may have a role to play in low-energy physics. 

\section*{Acknowledgments}

This work was supported in part by DOE Grant No.\ DE-FG02-95ER40893 and 
NATO Grand No.\ CRG-940784.

\end{document}